\RequirePackage{fix-cm}
\documentclass[natbib]{svjour3} 
\smartqed 
\usepackage{graphicx}
\usepackage{amsmath}
\usepackage{amssymb}
\usepackage{gensymb} 
\usepackage[utf8]{inputenc} 
\journalname{Celestial Mechanics and Dynamical Astronomy}
\begin{document}

\title{Non-resonant secular dynamics of trans-Neptunian objects perturbed by a distant super-Earth}
\titlerunning{Secular dynamics driven by a distant super-Earth} 

\author{Melaine~Saillenfest$^{1,2}$ \and Marc~Fouchard$^{1,3}$ \and Giacomo~Tommei$^{2}$ \and Giovanni~B.~Valsecchi$^{4,5}$}
\authorrunning{Melaine Saillenfest \and Marc Fouchard \and Giacomo Tommei \and Giovanni B. Valsecchi} 

\institute{
   M. Saillenfest \at
   melaine.saillenfest@obspm.fr\\
   \\
   $^1$ IMCCE, Observatoire de Paris, 77 av. Denfert-Rochereau, 75014 Paris, France\newline
   $^2$ Dipartimento di Matematica, Università di Pisa, Largo Bruno Pontecorvo 5, 56127 Pisa, Italia\newline
   $^3$ LAL-IMCCE, Université de Lille, 1 Impasse de l’Observatoire, 59000 Lille, France\newline
   $^4$ IAPS-INAF, via Fosso del Cavaliere 100, 00133 Roma, Italia\newline
   $^5$ IFAC-CNR, via Madonna del Piano 10, 50019 Sesto Fiorentino, Italia
}

\date{Received: 2017-8-3 / Accepted: 2017-6-27}

\maketitle

\begin{abstract}
   We use a secular model to describe the non-resonant dynamics of trans-Neptunian objects in the presence of an external ten-earth-mass perturber. The secular dynamics is analogous to an ``eccentric Kozai mechanism" but with both an inner component (the four giant planets) and an outer one (the eccentric distant perturber). By the means of Poincaré sections, the cases of a non-inclined or inclined outer planet are successively studied, making the connection with previous works. In the inclined case, the problem is reduced to two degrees of freedom by assuming a non-precessing argument of perihelion for the perturbing body.
   
   The size of the perturbation is typically ruled by the semi-major axis of the small body: we show that the classic integrable picture is still valid below about $70$~AU, but it is progressively destroyed when we get closer to the external perturber. In particular, for $a>150$~AU, large-amplitude orbital flips become possible, and for $a>200$~AU, the Kozai libration islands at $\omega=\pi/2$ and $3\pi/2$ are totally submerged by the chaotic sea. Numerous resonance relations are highlighted. The most large and persistent ones are associated to apsidal alignments or anti-alignments with the orbit of the distant perturber.
   
\end{abstract}

\section{Introduction}
   The hypothesis of a distant giant planet beyond Neptune is often proposed in the literature, as accounting for otherwise mysterious features of the Solar System. Indeed, it could reproduce the observed orbital clustering of the distant trans-Neptunian objects \citep{BATYGIN-BROWN_2016}, or the $6\degree$-tilt of the solar equator with respect to the planetary invariable plane \citep{BAILEY-etal_2016,GOMES-etal_2016}. In this paper, we will not discuss the possibility of existence of such a planet, but focus on the rich dynamical system it would create.
   
   With no distant perturber, the non-resonant secular dynamics beyond Neptune is well-known from \cite{GALLARDO-etal_2012} and \cite{SAILLENFEST-etal_2016a}. Using coplanar and circular orbits for the known planets, the equilibrium points and libration islands have well-determined locations and sizes in the space of orbital elements: for semi-major axes larger than $80$~AU, the only equilibrium points are located at $\omega=\pi/2$ and $3\pi/2$ for an inclination $I$ of about $63\degree$ or $117\degree$. The perihelion distance at equilibrium is obtained through the constant parameter $C_K=(1-e^2)\cos^2I$. Moreover, it has been shown that the maximum variations of the perihelion distance, given by the width of the libration islands, cannot exceed $16.4$~AU with that mechanism \citep[][]{SAILLENFEST-etal_2016a}. For realistic values of the eccentricities and inclinations of the giant planets (non-zero but small), numerical integrations of the secular system show that these structures are almost unaltered: the weak interaction with these new degrees of freedom makes the equilibrium points become periodic orbits, tightly bounded around their nominal values.
   
   In this paper, our goal is to study the effect of a distant perturber on the non-resonant secular dynamics of trans-Neptunian objects. The loss of symmetry implied by a massive body with significant eccentricity and inclination should disrupt this classical picture, so we aim at determining which features persist (if any) and how they transform under the perturbation.
   
   As pointed out by \cite{BATYGIN-BROWN_2016}, the output of numerical simulations get much more significance when understanding the underlying dynamical processes. Their analytical computations, though, were limited to very low-order terms, and strong assumptions were used to get integrable models. These limitations were observed by~\cite{BEUST_2016}, who subsequently studied the fully planar case. The corresponding secular system has only one degree of freedom so the dynamics is integrable: all the trajectories can be described by plotting the level curves of the secular Hamiltonian, with the semi-major axis as parameter. He reported secular equilibrium points with libration zones around $\Delta\varpi=0$ (at large perihelion distances) for non-crossing orbits, and $\Delta\varpi=\pi$ (at small perihelion distances) when the orbit of the particle crosses the trajectory of the distant planet. Then, he extended that model to introduce a mean-motion resonance between the particle and the distant planet, keeping a single degree of freedom by using the adiabatic approximation. The level curves of the resonant secular Hamiltonian function present a large variety of equilibrium points distributed at no particular value of $\Delta\varpi$. Of course, these results hold only for a completely planar problem. Besides, \cite{BEUST_2016} used only the second-order term of the inner planetary perturbation (although the outer planet component was fully computed). When considering an arbitrarily inclined small body, such a truncated model cannot give rise to the Lidov-Kozai mechanism, whereas it has been shown that it can have important effects for trans-Neptunian objects \citep{GALLARDO-etal_2012}.
   
   Secular models for hierarchical systems in the general spatial case appear in the literature with works as early as \cite{HARRINGTON_1968} for triple-star systems. Such models can be very efficient to capture the essence of the dynamics, so they are widely used and developed for systems with increasing complexity. We can mention for instance the recent work by \cite{HAMERS-etal_2015} describing the evolution of two planets orbiting a binary star, and its generalisation by \cite{HAMERS-PortegiesZwart_2016}. In the planetary case, and in particular when one body is massless, such models are the natural generalisation of the work of \cite{KOZAI_1962} so they are often said to raise an ``eccentric Kozai mechanism", with two degrees of freedom \citep[see for instance the review by][]{NAOZ_2016}. The non-zero eccentricity of the perturbing body makes possible a wide variety of trajectories, including striking orbital flips, during which the orbit of the small body switches suddenly from prograde to retrograde \citep[][]{KATZ-etal_2011,LITHWICK-NAOZ_2011,NAOZ-etal_2013,LI-etal_2014a}. In the context of the octupolar development of the secular system, \cite{LI-etal_2014b} studied the dynamics of a test-particle perturbed by an eccentric and inclined distant planet. Their exploration of the dynamics includes Poincaré sections with an approach similar to the one used in this paper. In our case, the presence of an inner axis-symmetric component should mix the features of both models: a \emph{classic} Kozai mechanism is produced by the known planets, while an \emph{eccentric} Kozai mechanism is driven by the distant super-Earth. In this article, we will restrict the study to orbits with a perihelion distance beyond Neptune, since they are much less chaotic and more likely to follow a secular dynamics. We will also focus on prograde orbits, since no retrograde object has been observed yet with a perihelion distance beyond Neptune. Finally, we recall that the models studied here are only valid in the absence of any mean-motion resonance, including chaotic diffusion driven by overlap of mean-motion resonances. Outside of such diffusive regions, the mean-motion resonances are quite confined in small intervals of semi-major axis, the largest ones being of the form $1\!:\!k$ \citep[see for instance][]{GALLARDO_2006b,GALLARDO_2006a}.
   
   In Sect.~\ref{sec:model}, we present the planetary model used and the secular Hamiltonian function. Then, Sect.~\ref{sec:planar} shows the exploration of the dynamics produced by a planar perturber on an arbitrarily inclined small body. That intermediate model makes the link between the studies of \cite{SAILLENFEST-etal_2016a} and \cite{BEUST_2016}, mixing up the properties of both systems. Finally, we give an approach of the general case in Sect.~\ref{sec:gen}, where we consider an eccentric and inclined distant planet but with a non-precessing argument of perihelion.
   
\section{Model and method}\label{sec:model}
   We use a set of $N$ inner planets evolving on circular and coplanar orbits, along with an eccentric and inclined outer planet.
   \begin{itemize}
      \item The inner component of the perturbation stands here for the currently known planets of the Solar System (which have indeed roughly circular and coplanar trajectories), with no distinction between terrestrial or giant planets. The circular-coplanar approximation is justified by the very small eccentricity and inclination of the giant planets of the Solar System, especially when measured in their invariable plane. More generally, such a model can be seen as the dominant term of an expansion in powers of the planetary eccentricities and inclinations \citep{THOMAS-MORBIDELLI_1996}.
      \item The mass of the outer planet is chosen to be ten earth-masses and its orbital elements at current epoch are given in Tab.~\ref{tab:orbel}. These values are within the best estimates obtained so far \citep{BROWN-BATYGIN_2016} and consistent with those used in the literature.
   \end{itemize}
   
   \begin{table}
      \centering
      \begin{tabular}{ c | c | c | c | c }
         $a'$ & $e'$ & $I'$ & $\omega'$ & $\Omega'$ \\
         \hline
         $700$~\text{AU} & $0.6$ & $30\degree$ & $150\degree$ & $113\degree$ \\
      \end{tabular}
      \caption{Current heliocentric orbital elements of the distant planet used in this paper. This ``nominal" orbit is used for instance by~\cite{FIENGA-etal_2016}.}
      \label{tab:orbel}
   \end{table}
   
   \noindent Such a system is qualitatively similar to those studied by \cite{INNANEN-etal_1997} or \cite{TAKEDA-etal_2008}, namely a tight planetary system orbited by a distant star companion. They showed that due mutual interactions, the precession of the inner system of planets under the action of an inclined distant perturber is “rigid” (the mutual inclinations remain small, as well as the eccentricities). In our case, the perturber is much less massive than a star companion, but \cite{BAILEY-etal_2016} and \cite{GOMES-etal_2016} showed that this mechanism could still be responsible for the tilt of the mean planetary plane of the Solar System with respect to the spin axis of the Sun.
   
   The reference plane used in this article coincides with the orbital plane of the $N$ inner planets, and the third axis is directed along their angular momentum. All the orbits are heliocentric. In the following, the orbital elements of the inner planets are written with the subscript $i$, whereas we use a prime for the outer planet.
   
   \subsection{Dynamics of the outer planet}
   Considering the distances involved, the inner planets are supposed negligibly affected by their distant companion. In the inclined case, we thus neglect the effect of rigid precession described above. Arguments favouring this simplification are given by the Roy-Walker parameters $\varepsilon^{23}$ and $\varepsilon_{32}$ \citep{WALKER-etal_1980}: whereas mutual parameters for the giant planets are of orders $10^{-4}$ to $10^{-6}$, the effects of the distant planet on the internal system range from $10^{-9}$ (Neptune) to $10^{-11}$ (Jupiter). Conversely, the effects of the giant planets on their distant companion are of order $10^{-7}$ to $10^{-8}$, so this is the next level of approximation to be taken into account, beyond strictly decoupled systems. Our simplification can also be justified using the work by \cite{TEYSSANDIER-etal_2013}: by rescaling or extrapolating their figures for the system considered here (typically Jupiter plus the distant ten-earth mass planet), we always end up in the ``white" region of their graphs, that is where nothing special happens.
   
   Hence, the long-term dynamics of the outer planet is accurately represented by a secular model as described by~\cite{GALLARDO-etal_2012} or \cite{SAILLENFEST-etal_2016a}. Since it is far from every equilibrium point of both non-resonant and resonant secular Hamiltonians, this planet is a typical case of ``decoupled" object with constant semi-major axis, eccentricity and inclination. Its long-term dynamics is thus accurately approximated by the leading-order term of the development in the semi-major axes ratios (the next term is indeed $10^4$ times smaller). Up to this level of approximation, the constant precession rates of $\omega'$ and $\Omega'$ are:
   \begin{equation}
      \left\{
      \begin{aligned}
         \dot{\omega}' &= \delta^2 \frac{3}{8}\big(5\cos^2I'-1\big) \\
         \dot{\Omega}' &= -\delta^2 \frac{3}{4}\cos I'
      \end{aligned}
      \right.
      \hspace{0.5cm}\text{with}\hspace{0.5cm}
      \delta^2 = \sqrt{\frac{a'}{\mu}}\left(\frac{1}{a'(1-e'^2)}\right)^2\sum_{i=1}^{N}\mu_i\left(\frac{a_i}{a'}\right)^2
   \end{equation}
   In these expressions, $a_i$ is the constant semi-major axis of the $i$th planet; $\mu$ and $\mu_i$ are the gravitational parameters of the Sun and of the $i$th planet, respectively. By including Jupiter, Saturn, Uranus and Neptune (the masses of the terrestrial planets being added to the Sun), we obtain the following numerical values:
   \begin{equation}\label{eq:precvalues}
      \dot{\omega}' = 0.201\,\text{rad}/\text{Gyr}
      \hspace{0.5cm}\text{and}\hspace{0.5cm}
      \dot{\Omega}' = -0.126\,\text{rad}/\text{Gyr}
   \end{equation}
   They can be verified by unaveraged numerical simulations. In the rest of the article, we will refer to these two quantities as $\nu_\omega'$ and $\nu_\Omega'$.
   
   At some points, we will also consider a planar outer planet, with orbital elements still given by Tab.~\ref{tab:orbel} but with $I'=0$. In that case, $\omega'$ and $\Omega'$ will be replaced by $\varpi'=\omega'+\Omega'$, with a precession rate of:
   \begin{equation}
      \dot{\varpi}' = \delta^2\frac{3}{4} \approx 0.146\,\text{rad}/\text{Gyr}
   \end{equation}
   We will refer to this last quantity as $\nu_\varpi'$ in the following\footnote{This is the same expression as the Eq.~2 by \cite{BATYGIN-BROWN_2016}, except that they give the associated period $2\pi/\nu_\varpi'$. Note that there is a typo error in their expression (the inverse of a sum is \emph{not} the sum of the inverses).}.
   
   \subsection{Osculating dynamics of the small body}
   We consider the orbit of a small body perturbed by both the $N$ inner planets and the precessing outer super-Earth. In Delaunay heliocentric elements, the corresponding Hamiltonian function is:
   \begin{multline}\label{eq:Hgen}
      \mathcal{H}\Big(\{\Lambda_i\},\Lambda',P_\omega',P_\Omega',L,G,H,\{\lambda_i\},\lambda',\omega',\Omega',\ell,g,h\Big) = \\
      \mathcal{H}_0\Big(\{\Lambda_i\},\Lambda',P_\omega',P_\Omega',L\Big) + \varepsilon\mathcal{H}_1\Big(L,G,H,\{\lambda_i\},\lambda',\omega',\Omega',\ell,g,h\Big)
   \end{multline}
   where the integrable part and the perturbation write respectively:
   \begin{equation}\label{eq:H}
      \left\{
      \begin{aligned}
         \mathcal{H}_0 &= -\frac{\mu^2}{2L^2} + \nu_\omega' P_\omega' + \nu_\Omega' P_\Omega' + \sum_{i=1}^Nn_i\,\Lambda_i + n'\,\Lambda'\\
         \varepsilon\mathcal{H}_1 &= - \sum_{i=1}^N\mu_i\left(\frac{1}{|\mathbf{r}-\mathbf{r}_i|} - \mathbf{r}\cdot\frac{\mathbf{r}_i}{|\mathbf{r}_i|^3}\right) - \mu'\left(\frac{1}{|\mathbf{r}-\mathbf{r}'|} - \mathbf{r}\cdot\frac{\mathbf{r}'}{|\mathbf{r}'|^3}\right)
      \end{aligned}
      \right.
   \end{equation}
   The vectors $\mathbf{r}$, $\mathbf{r}_i$ and $\mathbf{r}'$ are the heliocentric positions of the particle, of the $i$th inner planet, and of the outer one. The constants $\mu_i$ and $\mu'$ are the gravitational parameters of the planets, whereas $\{n_i\}$ and $n'$ are their mean motions. The momenta $\{\Lambda_i\}$ and $\Lambda'$ are conjugated to the mean longitudes $\{\lambda_i\}$ and $\lambda'$ of the planets, and $P_\omega'$ and $P_\Omega'$ are conjugated to $\omega'$ and $\Omega'$. They allow the definition of an autonomous system. We have then:
   \begin{equation}
      \left\{
      \begin{aligned}
         &\mathbf{r}_i \equiv \mathbf{r}_i(\lambda_i) \hspace{0.2cm}\text{for\ } i=1,2\dots N \\
         &\mathbf{r}' \equiv \mathbf{r}'(\lambda',\omega',\Omega') \\
         &\mathbf{r} \equiv \mathbf{r}(L,G,H,\ell,g,h)
      \end{aligned}
      \right.
   \end{equation}
   Finally, we recall that the Delaunay canonical coordinates $(L,G,H,\ell,g,h)$ are directly linked to the Keplerian elements $(a,e,I,\omega,\Omega,M)$ of the particle by:
   \begin{equation}
      \left\{
      \begin{aligned}
         L &= \sqrt{\mu a}\\
         G &= \sqrt{\mu a\,(1-e^2)}\\
         H &= \sqrt{\mu a\,(1-e^2)}\,\cos I
      \end{aligned}
      \right.
      \text{\ \ \ and\ \ }
      \left\{
      \begin{aligned}
         \ell &= M \\
         g &= \omega \\
         h &= \Omega
      \end{aligned}
      \right.
   \end{equation}
   The Hamiltonian system described by~\eqref{eq:Hgen} has $N+6$ degrees of freedom, but this number can be reduced using a geometric argument. Indeed, the perturbation involves $\omega'$ and $\Omega'$ only via the scalar product $\mathbf{r}\cdot\mathbf{r}'$ (this can be seen using a Legendre development of the inverse mutual distance). Computing that product in Keplerian elements and after some trigonometric manipulations, we get:
   \begin{equation}\label{eq:rr9}
      \begin{aligned}
         \frac{\mathbf{r}\cdot\mathbf{r}'}{r\,r'} &=  \sin(\alpha)\sin(\alpha')\sin(I)\sin(I') \\
      &  + \cos\big(\alpha -\alpha' + \Delta\Omega\big) \cos^2(I/2) \cos^2(I'/2) \\
      &  + \cos\big(\alpha +\alpha' + \Delta\Omega\big) \cos^2(I/2) \sin^2(I'/2) \\
      &  + \cos\big(\alpha +\alpha' - \Delta\Omega\big) \sin^2(I/2) \cos^2(I'/2) \\
      &  + \cos\big(\alpha -\alpha' - \Delta\Omega\big) \sin^2(I/2) \sin^2(I'/2)
      \end{aligned}
   \end{equation}
   where the norms $r\equiv|\mathbf{r}|$ and $r'\equiv|\mathbf{r}'|$ are independent of $\omega'$ and $\Omega'$. In that expression, the symbol $\alpha$ represents the sum of $\omega$ and the true anomaly (with a prime for the outer planet), and $\Delta\Omega=\Omega-\Omega'$. The longitudes of ascending nodes appear only via their difference, so it is possible to remove one degree of freedom by studying the system in a frame rotating with $\Omega'$. This is realised by a linear transformation involving the Delaunay angle $h=\Omega$:
   \begin{equation}
      \begin{pmatrix}
         \delta h \\
         \gamma
      \end{pmatrix}
      =
      \begin{pmatrix}
         1 & -1 \\
         0 &  1
      \end{pmatrix}
      \begin{pmatrix}
         h \\
         \Omega'
      \end{pmatrix}
   \end{equation}
   and applying its conjugated transposed on the momenta:
   \begin{equation}
      \begin{pmatrix}
         \tilde{H} \\
         \Gamma
      \end{pmatrix}
      =
      \begin{pmatrix}
         1 & 0 \\
         1 & 1
      \end{pmatrix}
      \begin{pmatrix}
         H \\
         P_\Omega'
      \end{pmatrix}
   \end{equation}
   This change of coordinates allows the momentum associated to $\delta h = \Delta\Omega$ to be simply $\tilde{H}=H$ so we will omit the ``tilde" sign in the following. In the new coordinates, the integrable part of the Hamiltonian function writes:
   \begin{equation}
      \mathcal{H}_0 = -\frac{\mu^2}{2L^2} + \nu_\omega' P_\omega' + \nu_\Omega' \Gamma - \nu_\Omega' H + \sum_{i=1}^Nn_i\,\Lambda_i + n'\,\Lambda'
   \end{equation}
   and the perturbation does not depend on $\gamma$. The momentum $\Gamma$ being a constant of motion, we will discard the term $\nu_\Omega' \Gamma$ in the Hamiltonian (thus dropping one degree of freedom).
   
   \subsection{Secular model}
   Assuming that the particle is far from any mean-motion resonance with the planets, we can get rid of the short-period angles by a close-to-identity change of coordinates. At first order of the perturbation, the Hamiltonian function in the new coordinates (hereafter named \emph{secular Hamiltonian}) is given by the average of $\mathcal{H}$ with respect to the fast independent angles $\ell$ and $\lambda_1,\lambda_2...\lambda_N,\lambda'$. Dropping the constant parts, it writes:
   \begin{equation}\label{eq:F}
      \mathcal{F}(P_\omega',L,G,H,\omega',g,\delta h) = \nu_\omega' P_\omega' - \nu_\Omega' H + \mathcal{F}_1(L,G,H,\omega',g,\delta h)
   \end{equation}
   where $\mathcal{F}_1$ is the numerically-computed average of $\varepsilon\mathcal{H}_1$. Even if we use the same symbols as before, we now manipulate the \emph{secular coordinates}. As usual for non-resonant secular models, the semi-major axis of the particle (momentum $L$) becomes a parameter.
   
   In the following, it is useful to have a normalized version of $\mathcal{F}$ which takes values of the order unity. This can be realised by adding a constant to $\mathcal{F}$ (dynamics unchanged) and multiplying it by a constant factor (change of the unit of time). For small bodies with a trajectory stretching between $a_N$ and the orbit of the outer planet, judicious values of these constants are given by the development in the semi-major axes ratios :
   \begin{equation}\label{eq:devNorm}
      \begin{aligned}
         \mathcal{F} &= \nu_\omega' P_\omega' - \nu_\Omega' H \\
         &- \frac{1}{a}\sum_{i=1}^{N}\mu_i - \frac{1}{a'}\mu'\\
         &- \frac{1}{a}\sum_{i=1}^{N}\mu_i\left(\frac{a_i}{a}\right)^2\frac{1}{8(1-e^2)^{3/2}}(3\cos^2I-1)\\
         &+ \mathcal{O}\left(\sum_{i=1}^{N}\mu_i\left(\frac{a_i}{a}\right)^4\right) + \mathcal{O}\left(\mu'\left(\frac{a}{a'}\right)^2\right)
      \end{aligned}
   \end{equation}
   where computational details can be found in \cite{LASKAR-BOUE_2010}, or \cite{SAILLENFEST-etal_2016a} for the inner component. The secular semi-major axis $a$ being a constant of motion, the normalised version of the secular Hamiltonian is chosen to:
   \begin{equation}
      \overline{\mathcal{F}} = \frac{\mathcal{F} - C_\text{offset}}{C_\text{scale}} 
   \end{equation}
   where the constant coefficients $C_\text{offset}$ and $C_\text{scale}$ are:
   \begin{equation}
      \left\{
      \begin{aligned}
         C_\text{offset} &= -\frac{1}{a}\sum_{i=1}^{N}\mu_i - \frac{1}{a'}\mu' \\
         C_\text{scale} &= \frac{1}{4\,a}\sum_{i=1}^{N}\mu_i\left(\frac{a_i}{a}\right)^2
      \end{aligned}
      \right.
   \end{equation}
   By this choice of scaling factor $C_\text{scale}$, we suppose that the second-order term of the development for the inner planets is the leading term of the Hamiltonian. This holds for small semi-major axes (a little beyond $a_N$), but not for large ones, for which the second order term of the development for the outer planet is more important. Moreover, the development \eqref{eq:devNorm} is valid only for trajectories entirely contained between the orbits of Neptune and of the outer planet, hence, the chosen coefficients have no clear dynamical meaning in the general case: they just allow to get a more ``human-readable" value for the Hamiltonian function (say, not too far from unity).
   
   The numerical computation of the secular Hamiltonian is now a common procedure in celestial mechanics. For one-degree-of-freedom secular systems, the Hamiltonian value with respect to the coordinates gives an immediate qualitative description of the dynamics, since every possible trajectory is defined by a distinct level curve. When two orbits cross, the resulting polar singularity of order 1 in the integral must be appropriately handled, but this is easily realised numerically and authors barely mention it anymore: \emph{the averaged Hamiltonian always exists, and is a continuous function, even in the planet-crossing case; this is because an improper integral over a two-dimensional torus of a function with a polar singularity of order 1 is absolutely convergent} \citep[quoted from][]{GRONCHI-MILANI_1998}. Hence, for one-degree-of-freedom secular systems, the geometry of the phase portraits with respect to the parameters is obtained in a plain way even for crossing orbits. Of course, as pointed out by \cite{THOMAS-MORBIDELLI_1996}, the secular approximation does not automatically hold when two osculating orbits cross, because of the possibility of actual physical collision (or very close encounter). However, each branch of the generalised secular trajectory is perfectly valid, so the latter give at least the geometry of the solutions in a piecewise way. In addition, \cite{GRONCHI-MILANI_1998} stress that particles with repeated orbit crossings can still exhibit very smooth behaviours on a secular timescale. The system studied by \cite{BEUST_2016} is even more critical, since it contains orbits which intersect \emph{at all time}. Nevertheless, he reported system lifetimes larger than the age of the Solar System (before the accidental occurrence of a dramatically close encounter which does invalidate the secular representation), showing the significance of the generalised secular model.
   
   The use of the numerically-computed secular system is less straightforward when there are several degrees of freedom: the complete equations of motion are required, and their calculation as well as their very mathematical definition are more problematic. Generically, any equation of motion can be obtained from $\mathcal{F}$ by inverting the partial derivative and the integral symbols: the chain rule is used from Cartesian, through Keplerian, to Delaunay coordinates, and the result is numerically averaged over the short-period angles. This amounts to consider the planets as massive interacting rings, forming what is called a ``N-ring system" by \cite{TOUMA-etal_2009}. The question to what extent this is equivalent to the time derivatives of the secular variables was extensively studied by \cite{GRONCHI-MILANI_1998} and \cite{GRONCHI_2002}. They demonstrated rigorously that this approach holds as long as the orbit of the small body does not cross any of the planetary orbits. During an orbital crossing, indeed, some of the partial derivatives are not defined (they present a polar singularity of order 2), even if the Hamiltonian itself is. However, they have a well-defined value arbitrarily close to the singularity on both sides, so in practice an orbital crossing results simply in a discontinuity of the ``force" term of the equations of motion. \cite{GRONCHI-MILANI_1998} showed that a generalised solution passing through the discontinuity can be uniquely defined as the trajectory connecting the limits of the incoming and outgoing smooth solutions\footnote{There is actually one very specific case where the generalised solution is not uniquely defined, namely when the crossing is exactly tangential. This can happen only if the mutual inclination of the asteroid and the planet is zero at the very moment of the orbital crossing. We will discard that case in this paper, since it has negligible probability to occur for an initially arbitrarily inclined small body.}. This generalised solution is necessarily non-smooth, but it is continuous. For a one-degree-of-freedom system, this corresponds to the usual level curves of the Hamiltonian, the crossings appearing as angular points. Note that an orbit crossing does not imply necessarily a collision in the non-averaged system, neither a chaotic behaviour (since the bodies can be located in very distant points of their orbits when they cross). \cite{GRONCHI-MILANI_2001} presented a practical algorithm to integrate numerically such a generalised secular trajectory: an integration step should never pass through the discontinuity, so the idea is to stop the integration exactly at the crossing point (limit of the \emph{left} smooth piece) and then restart it (\emph{right} smooth piece) without computing the force at the transition point. Some integrators, as those using a Runge-Kutta-Gauss scheme, do not need the calculation of the force at the initial nor the final points of a given step. Such an integrator must be used, at least for these two particular steps\footnote{\cite{GRONCHI-MILANI_2001} stress also the symplectic property of Runge-Kutta-Gauss integrators. The handling of the discontinuity, though, requires necessarily an adjustable integration step, which breaks the symplecticity of the overall scheme.}. In order to improve the stability of the numerical scheme even in the very neighbourhood of the transition, the discontinuous terms can be computed analytically using Kantorovich method with an appropriate intermediary function. For the sake of simplicity, we will not use it in this paper. In return, the conservation of the Hamiltonian will always be checked as a proxy of reliability of the numerical solutions (constancy of the normalized value at the $10^{-10}$ level). ``Bad" behaviours were found very rare and always avoidable by suitable integration steps.
   
   \subsection{Computational details}
   When two orbits cross, the calculation of an integration step arriving exactly on the transition point deserves some comments. \cite{GRONCHI-MILANI_2001} present an iterative procedure using nested dichotomy methods, associated with a polynomial extrapolation in order to detect \emph{a priori} when a crossing could occur. We preferred to use the method of \cite{HENON_1982} which seems to be more straightforward: when an integration sub-step is found to cross a discontinuity, the current step is immediately stopped and a unique, well-determined step is performed, arriving exactly on the desired point. This is made possible by a change of the independent variable used for the integration. In our case, two orbits cross when their mutual nodal distance vanishes, so the idea is to take this mutual nodal distance as a fictitious ``time" and to make a single step leading it to zero. The mutual nodal distance of the small body with an arbitrary planet $j$ (either the outer or an inner one) is given by:
   \begin{equation}\label{eq:Dnode}
      \Delta_j^\pm = \frac{a(1-e^2)}{1\pm e\cos\tilde{\omega}} - \frac{a_j(1-e_j^2)}{1\pm e_j\cos\tilde{\omega}_j}
   \end{equation}
   where $\pm$ stands for the ascending or descending mutual nodes. The angles $\tilde{\omega}$ and $\tilde{\omega}_j$ are the arguments of perihelion in the \emph{mutual} reference frame, defined by the $z$-axis being parallel to the angular momentum of the planet $j$ and the $x$-axis pointing toward the ascending mutual node of the small body (that is where its orbit crosses the $(x,y)$ plane from negative to positive $z$ values). We note that this reference frame is defined only for non-zero mutual inclinations, but it is anyway just a mathematical intermediate, used to define the orbit crossings. In terms of the Keplerian elements in the conventional reference frame, we get:
   \begin{equation}
      \begin{aligned}
         &\cos\tilde{\omega} = \frac{\cos\omega(\sin I\cos I_j - \cos I\sin I_j\cos\Delta\Omega_j) + \sin\omega\sin I_j\sin\Delta\Omega_j}{\sqrt{1-(\cos I\cos I_j + \sin I\sin I_j\cos\Delta\Omega_j)^2}}\\
         &\cos\tilde{\omega}_j = \frac{-\cos\omega_j(\sin I_j\cos I - \cos I_j\sin I\cos\Delta\Omega_j) + \sin\omega_j\sin I\sin\Delta\Omega_j}{\sqrt{1-(\cos I\cos I_j + \sin I\sin I_j\cos\Delta\Omega_j)^2}}
      \end{aligned}
   \end{equation}
   where $\Delta\Omega_j = \Omega-\Omega_j$. Naturally, the expression of $\Delta_j^\pm$ is greatly simplified for non-inclined planets on circular orbits (as the first $N$ planets considered in this paper). For any planet $j$, the time derivative of the mutual nodal distances can be computed in terms of our canonical coordinates, using the chain rule:
   \begin{equation}\label{eq:dDnodedt}
      \frac{\mathrm{d}\Delta_j^\pm }{\mathrm{d}t}= \frac{\partial \Delta_j^\pm}{\partial g}\dot{g} + \frac{\partial \Delta_j^\pm}{\partial \delta h}\dot{\delta h} + \frac{\partial \Delta_j^\pm}{\partial G}\dot{G} + \frac{\partial \Delta_j^\pm}{\partial H}\dot{H} + \frac{\partial \Delta_j^\pm}{\partial \omega'}\dot{\omega}'
   \end{equation}
   In practice, when a specific node crossing is detected, the corresponding mutual nodal distance \eqref{eq:Dnode} is taken as the new independent variable $\tau$. Noting generically $\rho$ its time derivative \eqref{eq:dDnodedt}, the new equations of motion are obtained by dividing the Hamilton equations by $\rho$. The evolution of the physical time $t$ must be added among the dynamical equations as:
   \begin{equation}
      \frac{\mathrm{d}t}{\mathrm{d}\tau} = \frac{1}{\rho}
   \end{equation}
   Hence, the exact position on the node crossing is obtained by a single integration step $\Delta\tau$ leading $\tau$ to zero. The conventional variables are then recovered to pursue the integration. The drawback of this method is that, when switching to the variable $\tau$, the integrator cannot make use of the previous integration steps\footnote{Correcting coefficients could actually be computed but they would require to save a lot of information from the previous steps.} (for instance to build a first guess for predictor-corrector iterations). In the same way, the restart of the integration in the conventional variables is equivalent to begin from scratch again. We considered, though, that the consequent increase of computation time was compensated by dropping the possibly numerous iterations otherwise required to reach the node crossing to machine precision.
   
   \subsection{Preliminary remarks}
   
   Before presenting our results, some comments about their comparison to previous works can be useful. As we mentioned in the introduction, the strategy used in this paper is similar to that of \cite{LI-etal_2014b}, and in both studies, the eccentric Kozai mechanism is raised by an outer planet acting on a test-particle. However, the comparison should be realised with care, for two major reasons:
   \begin{itemize}
      \item On the one hand, \cite{LI-etal_2014b}, as well as most works related to the eccentric Kozai mechanism, use a development of the Hamiltonian up to the octupolar term and the second order of the semi-major axis ratio. Of course, such a truncation is valid only for strictly hierarchical systems. They estimated this approximation to be valid for:
      \begin{equation}\label{eq:epsEKL}
         \frac{a}{a'}\frac{e'}{1-e'^2} < 0.1
      \end{equation}
      which amounts to $a < 75$~AU in our case, using the parameters in Tab.~\ref{tab:orbel}. In this article, this strong limitation is bypassed by using the full averaged Hamiltonian (obtained numerically), which is equivalent to a development containing an infinity of terms. We will consequently explore a parameter space which is well beyond the octupolar approximation.
      \item On the other hand, in our work the inner planets are the dominant part of the perturbation, especially for the small semi-major axes required in~\eqref{eq:epsEKL}. Even when the octupolar approximation could be valid, our results are thus strongly different from those obtained with models containing only the outer perturber.
   \end{itemize}
   In conclusion, the present study should be considered as an extension of the classic Kozai mechanism driven by inner planets to an additional external eccentric perturber, and not the contrary. Whereas numerous features of the eccentric Kozai mechanism are indeed revealed, they cannot be compared directly to previous works which use only the octupolar development\footnote{We use here a broader definition of ``classic" and ``eccentric" Kozai mechanisms than \cite{NAOZ-etal_2013} (right after their Eq.~26). Here, our definition holds for the non-truncated averaged Hamiltonian: it only indicates the orbit of the perturber, which is respectively circular or eccentric.}. This is why throughout this article, the comparison is mainly made with respect to the works by \cite{SAILLENFEST-etal_2016a}, in which the classic Kozai mechanism driven by the Solar System planets on an exterior test-particle is fully characterised.
   
   \section{Planar perturber}\label{sec:planar}
   By imposing the perturber inclination to be zero, the dependence on $\omega'$ and $\Omega'$ from~\eqref{eq:rr9} becomes:
   \begin{equation}
      \begin{aligned}
         \frac{\mathbf{r}\cdot\mathbf{r}'}{r\,r'} &= \cos\big(\alpha -v' + \Omega-\varpi'\big) \cos^2(I/2)\\
          &+ \cos\big(\alpha +v' - \Omega + \varpi'\big) \sin^2(I/2)
      \end{aligned}
   \end{equation}
   where $v'$ is the true anomaly of the distant planet. Since $\varpi$ is the only meaningful angle for zero-inclination orbits, the term $\nu_\omega' P_\omega' + \nu_\Omega' P_\Omega'$ in the osculating Hamiltonian~\eqref{eq:H} is replaced by $\nu_\varpi' P_\varpi'$. The variable $\varpi'$ acts just as $\Omega'$ from the general case, so the secular Hamiltonian for a planar perturber is simply:
   \begin{equation}\label{eq:Fplan}
      \mathcal{F}(L,G,H,g,\delta h) = - \nu_\varpi' H + \mathcal{F}_1(L,G,H,g,\delta h)
   \end{equation}
   where this time $\delta h = \Omega-\varpi'$. As before, the momentum $L$ (or equivalently the semi-major axis $a$ of the particle) is a free parameter. We are left with a two-degree-of-freedom system, non integrable in general, but which can be explored with Poincaré sections.
   
   A Poincaré section can be used for the mapping of a two-degree-of-freedom Hamiltonian system in a two-dimensional surface spanned by one pair of conjugated coordinates. This surface is defined by a fixed value of a function of the coordinates, as well as a direction of crossing. Besides, each map is parametrized by the value of the Hamiltonian. In practice, the computation of such a map consists in integrating numerically the equations of motion in a large range of initial conditions (with same Hamiltonian value), and retaining only the points where the obtained trajectories cross the section in the chosen direction. The method of \cite{HENON_1982} can be used once again, in order to get an integration point exactly on the surface. For two-degree-of-freedom systems, a Poincaré section allows to distinguish in a glance which trajectories are regular, as well as the size of the chaotic zones. Indeed, an integrable dynamics implies the existence of a second first integral (the first one being the Hamiltonian), so the corresponding trajectories evolve in a one-dimensional manifold. In practice, their section crossing points accumulate on continuous lines (quasi-periodic trajectories) or finite-numbered fixed points (periodic trajectories). On the contrary, chaotic trajectories evolve in a two-dimensional manifold, so their section crossing points are area-filling. Since a point of the map defines one and only one solution, a chaotic trajectory cannot cross the section inside a region filled with an integrable flow. Thus, authors often speak of ``stability islands embedded in a chaotic sea". This property implies the existence of ``stable chaos" \citep[so-called after][]{MILANI-NOBILI_1992}, for which chaotic trajectories are tightly trapped between two integrable manifolds. In that case, the corresponding chaotic zone looks more like a moat than an open sea.
   
   In all the following, the inner $N$ planets considered are Jupiter, Saturn, Uranus and Neptune ($N=4$), whereas the masses of the terrestrial planets are added to the Sun. The exploration of the parameter space is conducted as follows: for increasing values of the constant semi-major axis $a$, we present the most representative maps obtained when varying the value of the secular Hamiltonian $\overline{\mathcal{F}}$. The sections in both planes of conjugated coordinates are made simultaneously, so that we always present two maps for each value of $\overline{\mathcal{F}}$. In order to ease the interpretation, the momenta are replaced by non-canonical variables: we use the perihelion distance $q$ instead of $G$, and $H/L$ instead of $H$. Moreover, the ranges of inclination spanned by the represented trajectories are given along with the chosen values of $\overline{\mathcal{F}}$. Some examples of parameters for real objects are given in appendix~\ref{Asec:ic}. We use the following colour code for the points on the maps:
   \begin{itemize}
      \item[$\bullet$] \emph{Black} -- for integrable non-resonant trajectories. A fixed point on the maps corresponds to oscillations of the angle itself (by opposition to a resonant combination).
      \item[$\bullet$] \emph{Blue} -- for integrable trajectories driven by a resonance between the two degrees of freedom. A fixed point on the maps corresponds to oscillations of a linear combination of the two angles (which individually circulate\footnote{Such a simple colour code can be a bit ambiguous, in particular for resonances between the oscillation frequency of one angle and the circulation frequency of the other: they are represented in blue even if one angle oscillates. This should not mislead the reader, though, since further indications are given in the captions and in the text.}). Among them, \emph{large green dots} are used to draw the $1\!:\!1$ resonances, in order to help the reader to distinguish the different features.
      \item[$\bullet$] \emph{Red} -- for chaotic trajectories, that is with unpredictable crossing points on the section spreading in a surface. This surface can be very large, or tightly packed between integrable curves.
   \end{itemize}
   Finally, note that some regions of the maps are forbidden by the chosen value of the Hamiltonian: in our figures, such regions are represented in grey.
   
   With no distant perturber, the equilibrium points and corresponding libration islands are well-known from~\cite{GALLARDO-etal_2012} and \cite{SAILLENFEST-etal_2016a}. If these equilibrium points persist in the perturbed problem, they are expected to become periodic orbits (mapped in discrete points on the sections), surrounded by quasi-periodic trajectories (mapped as curves).
   
\begin{figure}
      \centering
      \includegraphics[width=\textwidth]{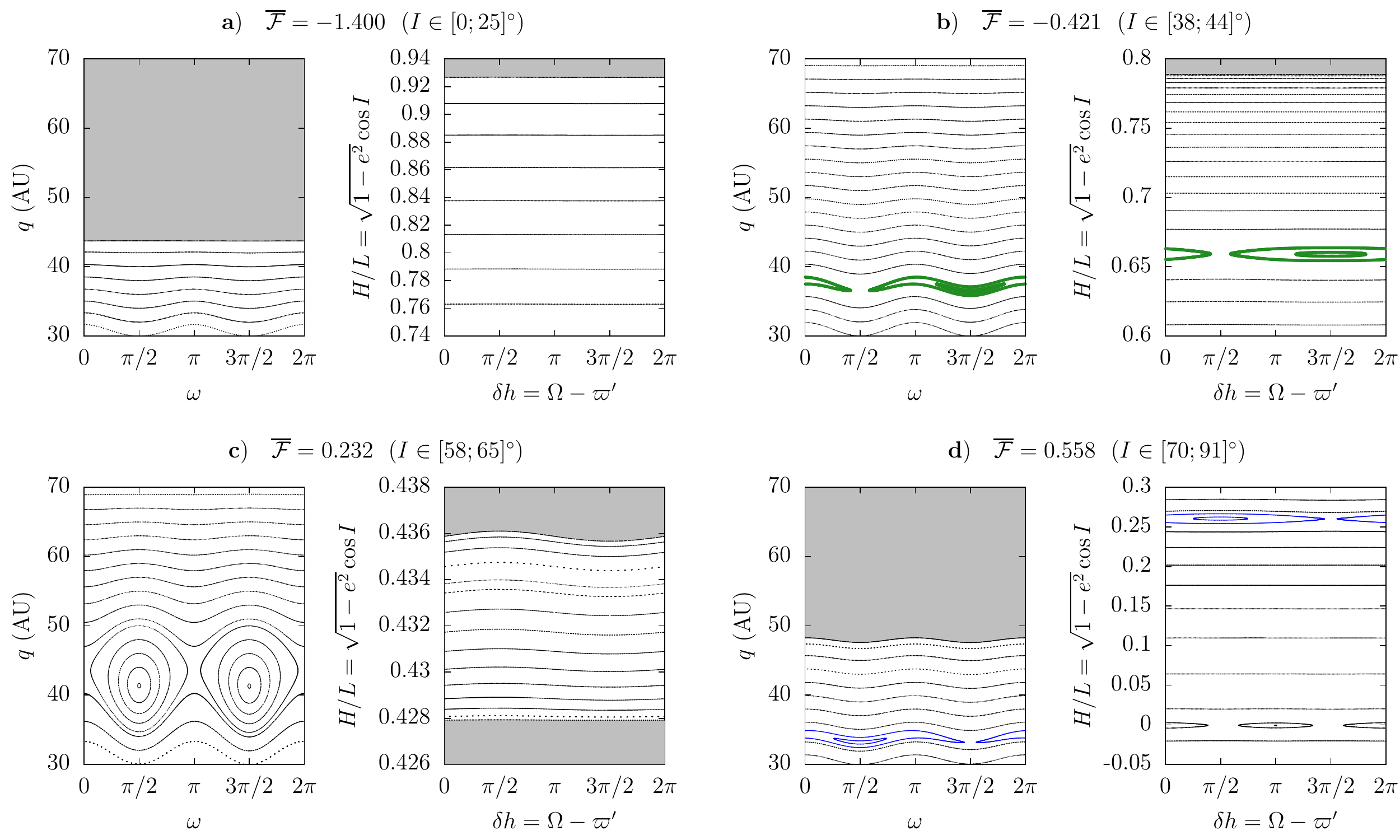}
      \caption{Poincaré maps for a planar perturber. The constant semi-major axis of the particle is $a=70$~AU. Each of the four panels (\textbf{a},\textbf{b},\textbf{c},\textbf{d}) corresponds to a different value of the secular Hamiltonian. The range of inclination given for each panel is the range spanned by all the represented trajectories. Every section in the $(\omega,q)$ plane is made for $\delta h=\pi/2$ and $\dot{\delta h}<0$. Every section in the $(\delta h,H)$ plane is made for $\omega=\pi/2$, with $\dot{\omega}>0$ for (\textbf{a},\textbf{b}) and $\dot{\omega}<0$ for (\textbf{c},\textbf{d}). The panels \textbf{b} and \textbf{d} feature the resonances $\omega+\delta h$ and $\omega-\delta h$, respectively. The panel \textbf{c} shows two fixed points for $\omega$ (at $I\approx 63\degree$), and the panel \textbf{d} shows two fixed points for $\delta h$ (at $I\approx 90\degree$).}
      \label{fig:P9plan_70}
   \end{figure}

   \bigskip
   Figure~\ref{fig:P9plan_70}, computed for $a=70$~AU shows that the effect of the distant planet is almost unnoticeable for small semi-major axes. Indeed, the most notable features of the maps are driven by the inner planets: the classic equilibrium points at $\omega$ equal to $\pi/2$ and $3\pi/2$ are easily recognisable and the quantity $\sqrt{1-e^2}\cos I$ is almost conserved. In that particular case, the maps are very close to the trajectories in the physical space itself (which oscillates slightly around the lines on the sections). The only extra features due to the distant planet have a very little impact on the dynamics. They are namely:
   \begin{enumerate}
      \item[a)] The libration zones around $\omega$ equal to $\pi/2$ and $3\pi/2$ allow slightly larger oscillations of the perihelion distance.
      \item[b)] The resonances $1\!:\!\pm 1$ appear between the two angles (with respective resonant angles $\omega+\delta h$ and $\omega-\delta h$), but they have a very little effect on the dynamics.
      \item[c)] The degeneracy of the $I=90\degree$ line is removed: it splits in two fixed points at $\delta h=0$ and $\pi$ surrounded by thin libration islands. We recall that without eccentric perturber, the $H=0$ line is entirely composed of equilibrium points for $\Omega$.
   \end{enumerate}
   Note that the classic equilibrium points of $\omega$ at about $I=63\degree$ divide the zones where $\omega$ circulates towards the right ($\dot{\omega}>0$ below the islands) from the zones where it circulates towards the left ($\dot{\omega}<0$ above the islands). In the same way, the equilibrium points of $\delta h$ at about $I=90\degree$ divide the zones where $\delta h$ circulates towards the right ($\dot{\delta h}>0$ for $I>90\degree$) from the zones where it circulates towards the left ($\dot{\delta h}<0$ for $I<90\degree$). It is important to keep it in mind all along this paper, since the Poincaré sections are defined for a specific direction of crossing. Here, we mainly focus on prograde orbits, thus with $\dot{\delta h}<0$, however, some sections feature also several trajectories with slightly negative momentum $H$, which consequently do not produce any point on the $(\omega,q)$ maps.
   
   The confrontation with the results of \cite{BEUST_2016} deserves some comments. In the fully planar case, $\omega$ and $\Omega$ are replaced by $\varpi$, and his figures are plotted in the $(\Delta\varpi,e)$ plane. Using our set of coordinates, $\Delta\varpi$ writes $\omega+\delta h$, so the equilibrium points reported by \cite{BEUST_2016}, corresponding to apsidal alignment or anti-alignment, are equivalent to the resonance $1\!:\!1$ (drawn in green) in our more general model\footnote{For a more straightforward comparison with \cite{BEUST_2016}, we could have taken directly the angle $\Delta\varpi=\omega+\delta h$ as canonical coordinate. However, the other resonances would have become harder to interpret (for instance $\omega-\delta h$ turns to $2\omega-\Delta\varpi$), and we would have lost the property of the equilibrium points of $\omega$ and $\delta h$, dividing prograde from retrograde resonances.}. This has a direct consequence: in the weakly perturbed system, apsidal confinement is only possible when $\omega$ and $\delta h$ circulate in opposite directions, that is for inclinations below $63\degree$, or lying between $90\degree$ and $117\degree$. A direct link to the study of \cite{BEUST_2016} is also given by the limits of the forbidden (grey) regions on the panels compatible with $I=0$. These limits are precisely given by $I=0$, so they represent a specific level curve of the planar Hamiltonian function. This is not very informative for Fig.~\ref{fig:P9plan_70} since the limit is very flat (panel \textbf{a}). In the next figures, however, the positions and shapes of the libration zones will be clearly recognisable. To ease the comparison, we added in appendix the level curves of the Hamiltonian in the completely planar case (Fig.~\ref{fig:Hsec-plan}). Each of these level curves could represent the limit of a forbidden region in a Poincaré section for the spatial case.
   
   \cite{BEUST_2016} reported no stable equilibrium point for small semi-major axes: there is actually one at $\Delta\varpi=0$ (corresponding to an apsidal alignment with the distant planet), but located at non-zero inclinations. This can be seen on the panel \textbf{b} of Fig.~\ref{fig:P9plan_70}, where the centre of the resonant trajectories corresponds to $\Delta\varpi=0$, whereas the $\Delta\varpi=\pi$ point lies on the separatrix\footnote{These sections are made respectively for $\delta h$ and $\omega$ equal to $\pi/2$, so a fixed point at $3\pi/2$ means for both sections an equilibrium point of $\Delta\varpi=\omega+\delta h$ at $0$.}. Note that $\Delta\varpi$ oscillates but $\omega$ and $\delta h$ circulate in opposite directions.
   
\begin{figure}
      \centering
      \includegraphics[width=\textwidth]{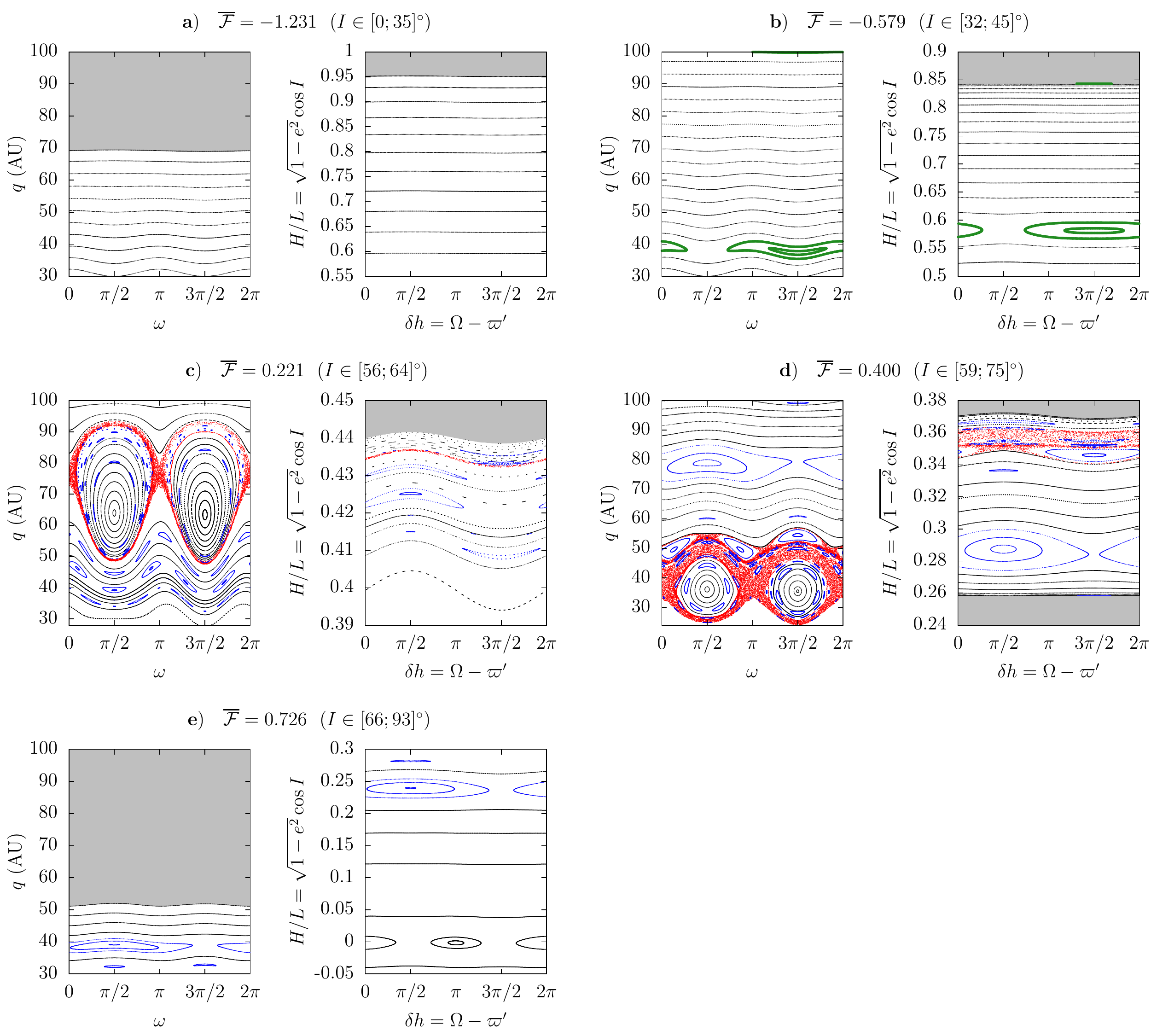}
      \caption{Poincaré maps for a planar perturber (see text for details). The constant semi-major axis of the particle is $a=100$~AU. Every section in the $(\omega,q)$ plane is made for $\delta h=\pi/2$ and $\dot{\delta h}<0$. Every section in the $(\delta h,H)$ plane is made for $\omega=\pi/2$, with $\dot{\omega}>0$ for (\textbf{a},\textbf{b},\textbf{c}) and $\dot{\omega}<0$ for (\textbf{d},\textbf{e}). Among others, the panels \textbf{b} and \textbf{e} feature the resonances $\omega+\delta h$ and $\omega-\delta h$ (same as in Fig.~\ref{fig:P9plan_70}). On the panels \textbf{c} and \textbf{d} numerous resonances are embedded in the chaotic sea (the corresponding resonant angles are given by the number of islands on the left and right graphs). The blue islands organised around the fixed points of $\omega$ are resonances between the libration frequency of $\omega$ and the circulation frequency of $\delta h$. A zoom-in view reveals secondary resonances as well.}
      \label{fig:P9plan_100}
   \end{figure}

   \bigskip
   For $a=100$~AU, Fig.~\ref{fig:P9plan_100} shows that a chaotic zone shows up around the fixed points of $\omega$ (panels \textbf{c} and \textbf{d}). The libration islands for $\omega$ are besides very enlarged with respect to their maximum width of $16.4$~AU without distant perturber. Various resonances appear between the two angles, including resonances between the circulation frequencies of $\omega$ and $\delta h$, resonances between the libration frequency of $\omega$ and the circulation frequency of $\delta h$, as well as secondary resonances. As usual for Poincaré sections, the maps show only the most obvious ones: a more careful analysis would reveal a lot of complex high-order resonances hidden in the chaos. On the panel \textbf{b} of Fig.~\ref{fig:P9plan_100}, there is a very thin island of apsidal alignment ($\Delta\varpi=0$) at high perihelion distances (barely noticeable). Contrary to other resonances present in Figs.~\ref{fig:P9plan_70} and \ref{fig:P9plan_100}, which can adopt various perihelion distances when varying slightly the Hamiltonian value, its position is fixed: it always remains close to the circular orbit. This is the precursor of the $\Delta\varpi=0$ equilibrium point reported by~\cite{BEUST_2016} in the planar case. For such a small semi-major axis, it is though limited to non-zero inclinations and a narrow range of Hamiltonian values.
   
\begin{figure}
      \centering
      \includegraphics[width=\textwidth]{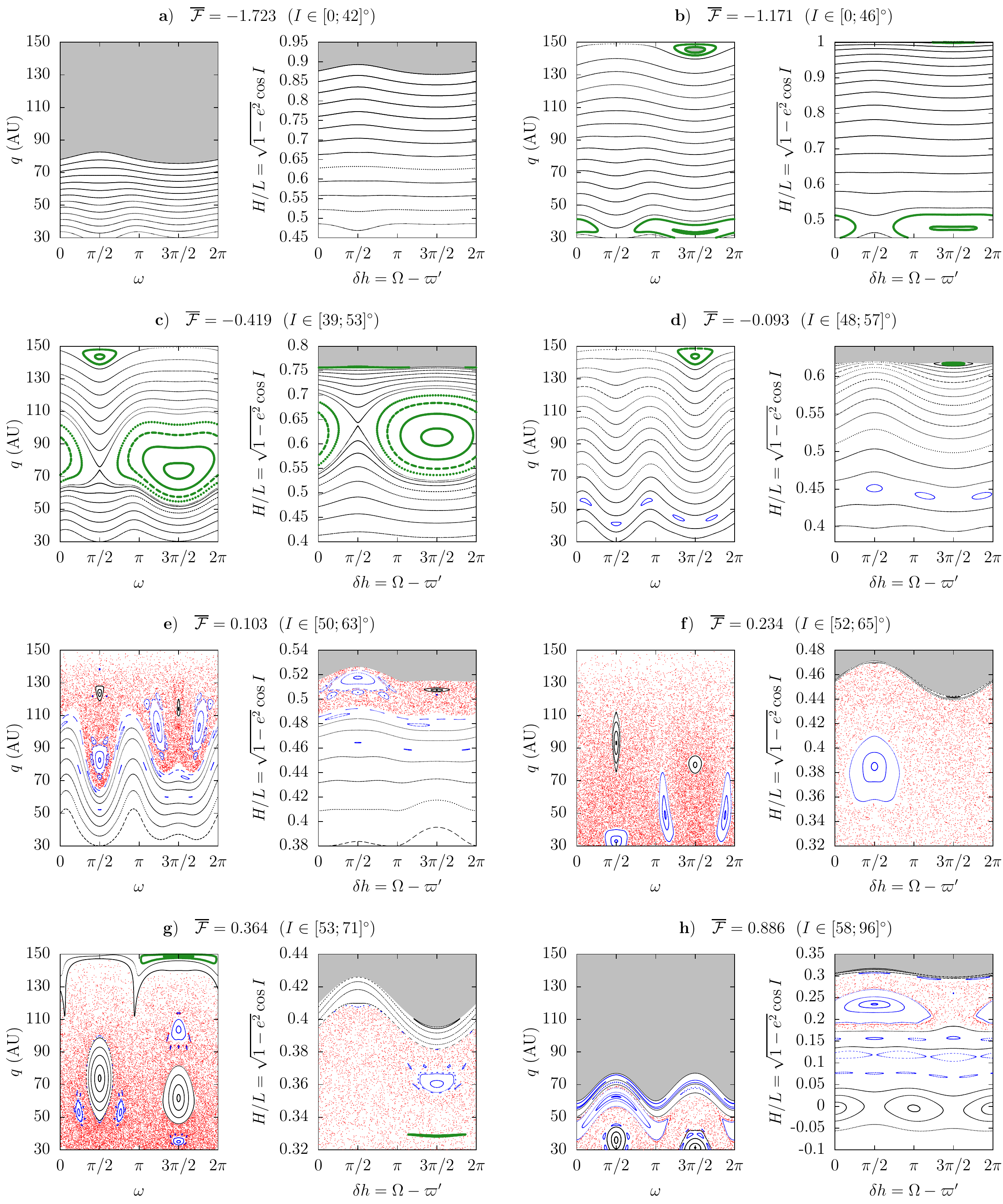}
      \caption{Poincaré maps for a planar perturber. The constant semi-major axis of the particle is $a=150$~AU. Every section in the $(\omega,q)$ plane is made for $\delta h=\pi/2$ and $\dot{\delta h}<0$. Every section in the $(\delta h,H)$ plane is made for $\omega=\pi/2$, with $\dot{\omega}>0$ for (\textbf{a}--\textbf{g}) and $\dot{\omega}<0$ for \textbf{h}. On the panel \textbf{e}, the libration island of $\omega$ around $\pi/2$ appears only as a $1\!:\!1$ resonance between the libration frequency of $\omega$ and the circulation frequency of $\delta h$ (small black island at $\delta h=3\pi/2$). We used black lines, though, to stress that $\omega$ itself oscillates around $\pi/2$, but note that $\delta h$ actually circulates, despite the closed black curves drawn on the right.}
      \label{fig:P9plan_150}
   \end{figure}

   \bigskip
   For $a=150$~AU, that equilibrium point is much more obvious (panels \textbf{b} and \textbf{d} of Fig.~\ref{fig:P9plan_150}). We added extra panels to detail its evolution with inclination. On the panels \textbf{a} and \textbf{b}, remember that the limits of the grey zones correspond to the zero inclination case. Knowing the position of the planar equilibrium points from Fig.~\ref{fig:Hsec-plan}, we can determine whether they persist or not for inclined orbits. Indeed, the upper green points on the panel \textbf{b} come from very slightly inclined trajectories, and they enclose completely the zero-inclination limit (the small grey zone detached from the top). This implies that the $\Delta\varpi=0$ equilibrium is transported continuously toward non-zero inclinations. For more inclined orbits, the fixed point switches from apsidal alignment to anti-alignment in a small range of inclinations (panel \textbf{c} near the circular orbit). Moreover, the other $1\!:\!1$ resonance, already present for smaller semi-major axes, now stretches in a much wider region of the phase space (panel \textbf{c}), multiplying the possibilities of $\Delta\varpi$ oscillations. Hence, the claim of \cite{BEUST_2016} that the non-resonant dynamics is able to produce both apsidal alignment and anti-alignment is widely generalised for inclined bodies. For semi-major axes as modest as $a=150$~AU, though, the aligned case is clearly favoured. On the overall Fig.~\ref{fig:P9plan_150}, we see that the possible excursion in inclination for a fixed Hamiltonian value is much wider than for smaller semi-major axes, making appear the classic $\omega$ fixed points on more numerous panels (\textbf{e}--\textbf{h}). However, the libration islands are quite ``nibbled" by the surrounding chaotic sea, so they appear much thinner than in Fig.~\ref{fig:P9plan_100}. The asymmetry of the two islands is due to the fixed value of $\delta h$ used to build the map: the geometry of the two islands is inversed by taking $3\pi/2$ instead of $\pi/2$ (mirror symmetry). This dependence of the chosen section plane is another indicator of the stronger interaction between the two degrees of freedom.
   
\begin{figure}
      \centering
      \includegraphics[width=\textwidth]{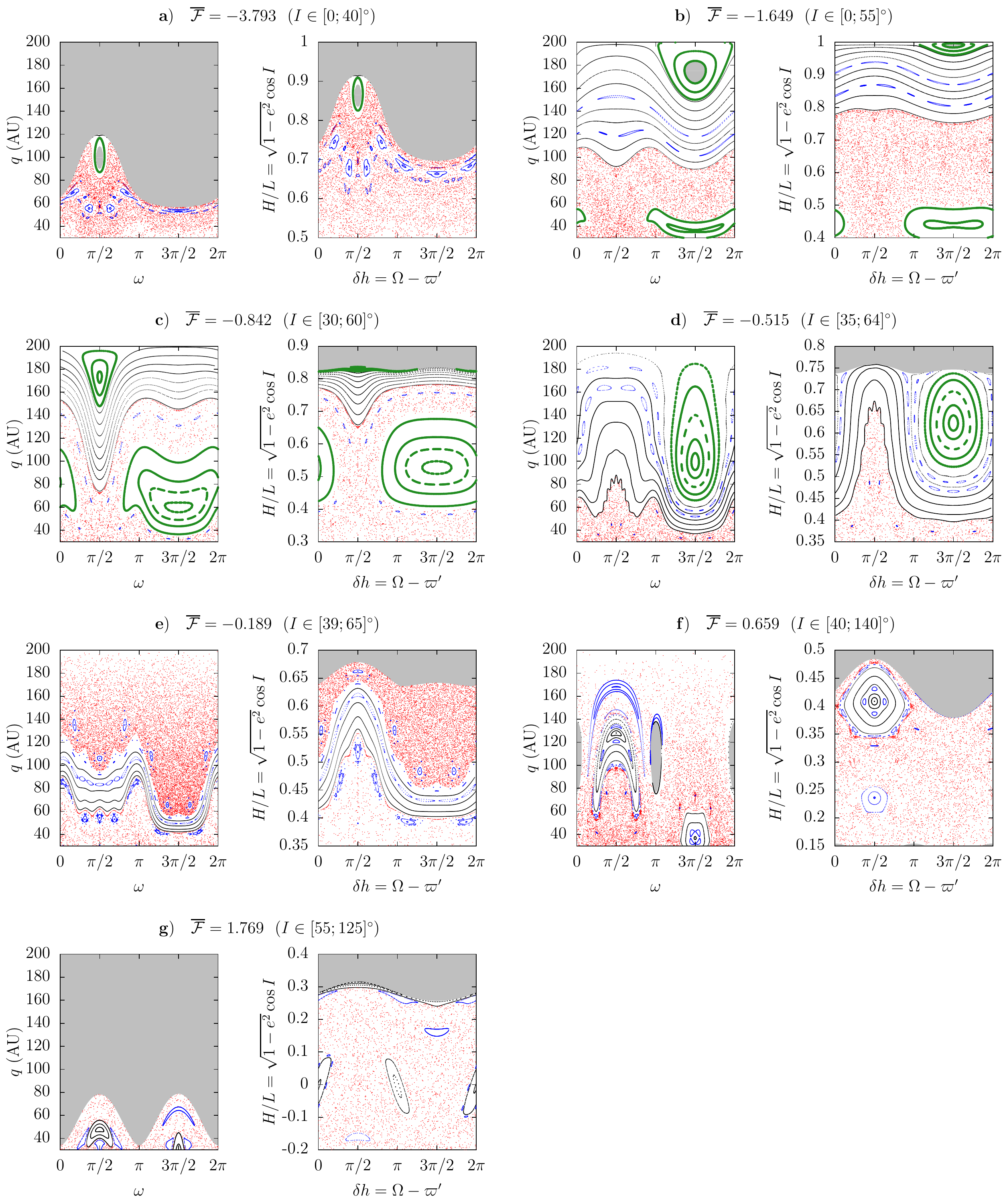}
      \caption{Poincaré maps for a planar perturber. The constant semi-major axis of the particle is $a=200$~AU. Every section in the $(\omega,q)$ plane is made for $\delta h=\pi/2$ and $\dot{\delta h}<0$. Every section in the $(\delta h,H)$ plane is made for $\omega=\pi/2$, with $\dot{\omega}>0$ for (\textbf{a}-\textbf{e}) and $\dot{\omega}<0$ for (\textbf{f},\textbf{g}). On the panel \textbf{f}, the libration island of $\omega$ around $\pi/2$ appears only as a $1\!:\!1$ resonance between the circulation frequency of $\delta h$ and the libration frequency of $\omega$. We used black lines, though, to stress that $\omega$ itself oscillates around $\pi/2$ (but note that $\delta h$ actually circulates, despite the closed black curves drawn on the right). On the panel \textbf{e}, the different density of red points between the two chaotic zones has a purely numerical origin: orbits in the lowermost region are subject to repeated close encounters with either the outer or the internal planets, slowing down the computations. As a consequence, only a few points are obtained during a reasonable computing time.}
      \label{fig:P9plan_200}
   \end{figure}

   \bigskip
   For $a=200$~AU (Fig.~\ref{fig:P9plan_200}), the $\Delta\varpi$ libration island near the circular orbit becomes wider as the fixed point moves toward higher eccentricities (panels \textbf{b},\textbf{c},\textbf{d}). The evolution of its position and shape for varying $a$ is thus generalised to non-zero inclinations. As before, it becomes a $\Delta\varpi = \pi$ libration island in some range of inclination (panel \textbf{c}). The other $\Delta\varpi = 0$ island, on the contrary, which was very large for $a=150$~AU, is now surrounded by a chaotic zone (panels \textbf{b} and \textbf{c}). In some range of Hamiltonian values, it merges with the upper $1\!:\!1$ resonance and produces a very wide island (panel \textbf{d}). For slightly higher values of the Hamiltonian, though, that island turns to a chaotic zone (panel \textbf{e}), which announces the proximity of the classic equilibrium points for $\omega$. Note that numerous orbits now intersect the trajectory of Neptune and/or of the distant planet, especially in the chaotic regions. For trajectories with $I=0$, we know from \cite{BEUST_2016} that the intersecting orbits produce an equilibrium point at $\Delta\varpi=\pi$ (apsidal anti-alignment). This results in the detached grey zone in the panel \textbf{a}. It is surrounded by a thin quasi-periodic flow, showing that the libration island persists for very small inclinations (the green curve represented oscillates between $I=0.1\degree$ and $0.5\degree$). For more inclined orbits, the chaos dominates but still sticking around the resonance. The chaos spreads also around the $\delta h$ equilibrium points (panel \textbf{g}), allowing chaotic orbital flips between prograde and retrograde orbits. This is quite different from the regular orbits oscillating around $I=90\degree$ (present also for smaller semi-major axes), since this time the orbit can stay retrograde for a long period of time, according to its wandering inside the chaotic zone. Very inclined and retrograde objects are actually observed in the distant Solar System\footnote{On 2017-06-06, the JPL Small-Body Database Search Engine reports 8 non-cometary objects with $a>150$~AU, $q>5$~AU and $I>50\degree$ (\texttt{https://ssd.jpl.nasa.gov/sbdb\_query.cgi}).}, and their formation was studied in particular by \cite{GOMES-etal_2015} and \cite{BATYGIN-BROWN_2016b}. The latter pointed out that highly-inclined objects with small semi-major axes can still be explained by this mechanism, through a subsequent diffusion of semi-major axis due to the inner giant planets. Of course, this last effect cannot appear in a secular model as ours. Finally, the most striking features in Fig.~\ref{fig:P9plan_200} are the two forbidden regions at $\omega=0$ and $\pi$ on the panel \textbf{f}. They correspond to oscillations of \emph{both} $\omega$ and $\delta h$ around $0$ or $\pi$, thus producing no points on the sections. In these regions, $\Delta\varpi$ oscillates also around $0$ or $\pi$, leading to ``frozen" aligned or anti-aligned orbits. The inclination of these trajectories oscillates around $90\degree$ (which is impossible to see in Fig.~\ref{fig:P9plan_200} because of the parameters chosen for the section), leading to a very particular geometry avoiding close orbital approaches. That kind of orbit is described more in detail below.
   
\begin{figure}
      \centering
      \includegraphics[width=\textwidth]{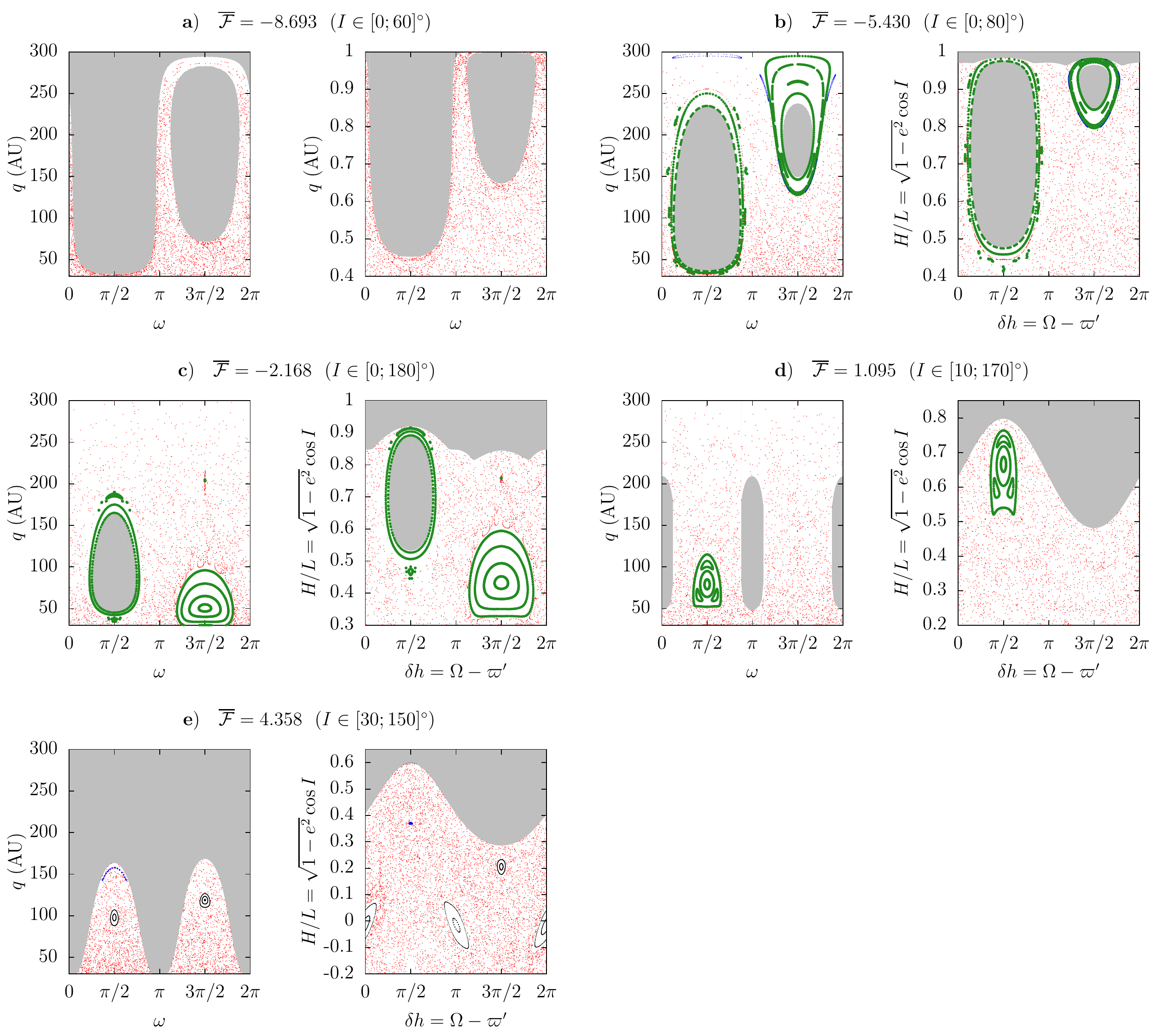}
      \caption{Poincaré maps for a planar perturber. The constant semi-major axis of the particle is $a=300$~AU. Every section in the $(\omega,q)$ plane is made for $\delta h=\pi/2$ and $\dot{\delta h}<0$. Every section in the $(\delta h,H)$ plane is made for $\omega=\pi/2$, with $\dot{\omega}>0$ for (\textbf{a}-\textbf{d}) and $\dot{\omega}<0$ for \textbf{e}. The axes ranges are mainly focussed on prograde orbits ($H>0$), but most of the chaotic orbits can flip (see the inclination ranges). On the panel \textbf{e}, the libration island of $\omega$ around $\pi/2$ appears only as a $1\!:\!1$ resonance between the circulation frequency of $\delta h$ and the libration frequency of $\omega$ (black island at $\delta h=3\pi/2$). We used black lines, though, to stress that $\omega$ itself oscillates around $\pi/2$ (but note that $\delta h$ actually circulates, despite the closed black curves drawn on the right). On the panel \textbf{d} the two forbidden regions at $\omega=0$ and $\pi$ correspond to oscillations of both $\omega$ and $\delta h$ around $0$ or $\pi$, thus producing no points on these sections.}
      \label{fig:P9plan_300}
   \end{figure}

\begin{figure}
      \centering
      \includegraphics[width=\textwidth]{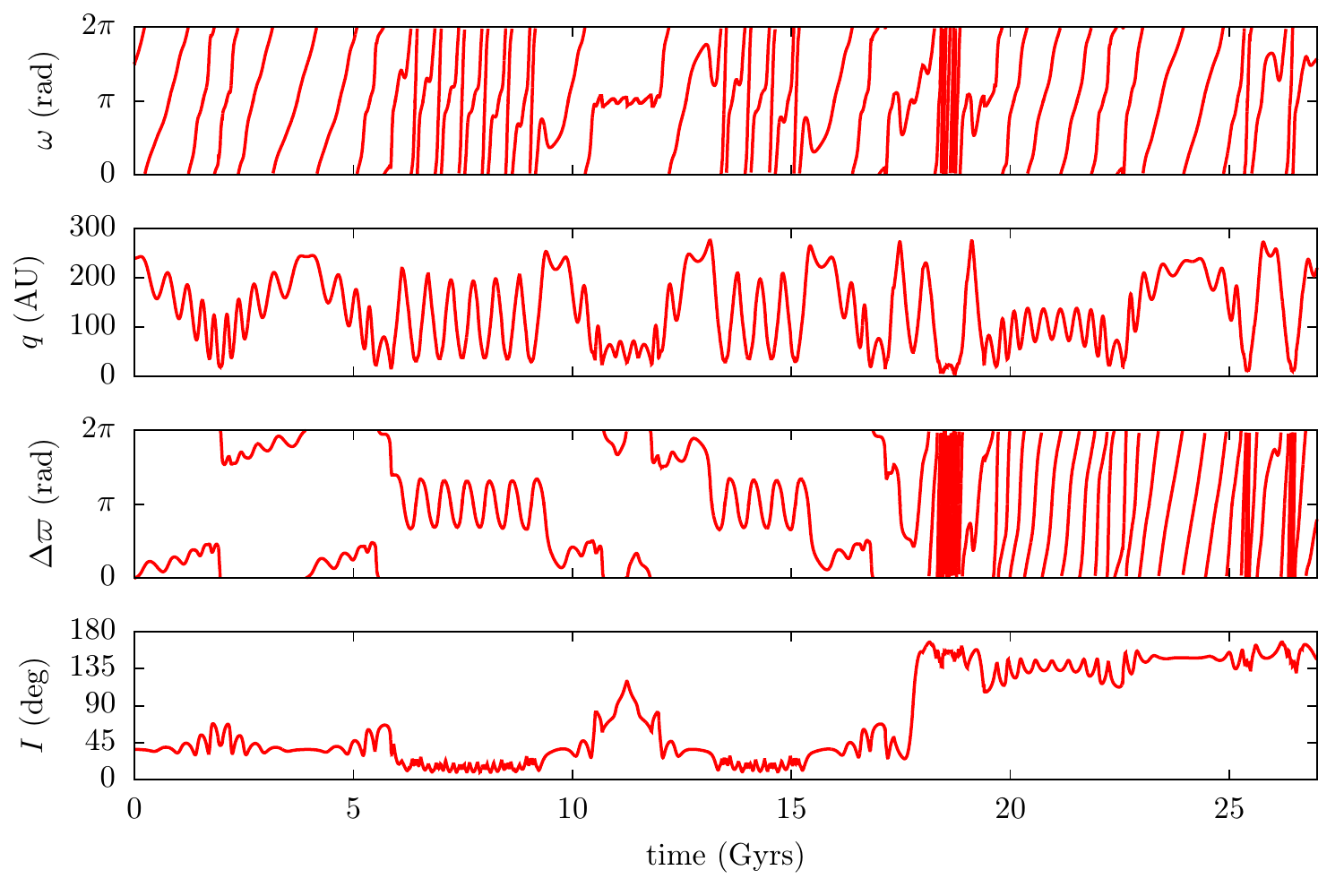}
      \caption{Complete evolution of a chaotic trajectory with $a=300$~AU. It contributes to the red dots on the maps \textbf{c} in Fig.~\ref{fig:P9plan_300}. The initial conditions are $(\omega,\delta h) = (3\pi/2,\pi/2)$ and $(q,H/L) = (240~\text{AU},0.784530340279072)$. Firstly, the particle switches from apsidal alignment to anti-alignment (both with circulating $\omega$ and $\delta h$). Then it adopts a frozen aligned orbit ($\omega\approx\delta h\approx\pi$) with $I$ oscillating around $90\degree$, and after brief states in apsidal anti-alignment and alignment, it flips to a retrograde orbit. In the retrograde state, it sticks to the $\omega-\delta h$ resonance, symmetric to the prograde $\omega+\delta h$ one, but producing no particular alignment in the physical space. Note that the apsidal anti-alignments are realised at very small inclinations, since the corresponding points on the Poincaré section lie near the limit of the forbidden zone (Fig.~\ref{fig:P9plan_300}\textbf{c}), around the few regular trajectories.}
      \label{fig:detail_300}
   \end{figure}

   \bigskip
   For $a=300$~AU, the only substantial stable features consists in resonances of apsidal alignment and anti-alignment (Fig.~\ref{fig:P9plan_300}). The previous equilibrium points of $\omega$ (at $\pi/2$ and $3\pi/2$) and of $\delta h$ (at $0$ and $\pi$) persist only marginally on the panel \textbf{e}. Emerging from $I=0$, the stable anti-aligned trajectories reach now moderate inclinations. For instance the resonant orbits on the panel \textbf{d} evolve between $I=10\degree$ and $25\degree$. On the panel \textbf{c}, note the presence of the two kinds of apsidal alignment: the very tiny green orbit at $q\approx 200$~AU is the residual of the island emerging from $I=0$, whereas the bottom one is the usual $1\!:\!1$ resonance present in every previous figure. Finally, the two forbidden zones on the panel \textbf{d} at $\omega=0$ and $\pi$ correspond also to apsidal alignments: they are filled with frozen orbits with both $\omega$ and $\delta h$ oscillating around $0$ or $\pi$, whereas $I$ oscillates around $90\degree$ (same as for $a=200$~AU, Fig.~\ref{fig:P9plan_200}\textbf{f}). The variations in inclination allowed in the chaotic zones are very large, and orbital flips are allowed in almost every panel. Note that the retrograde region of the phase space (not shown) is identical to the prograde one (mirror symmetry), with $\delta h$ circulating in the opposite direction. Hence, the particles on chaotic trajectories can jump indifferently from prograde resonances ($\omega+\delta h$) to retrograde ones ($\omega-\delta h$). In the retrograde case, though, the resonance does not correspond to a particular orbital alignment. The complete orbital evolutions of the chaotic trajectories reveal transient states in every regime presented in Fig.~\ref{fig:P9plan_300}, with sticky chaos. Fig.~\ref{fig:detail_300} presents a typical example, contributing to fill with red dots the chaotic sea of Fig.~\ref{fig:P9plan_300}\textbf{c}. We recognise apsidal alignments, apsidal anti-alignments, aligned frozen orbits with $I=90\degree$, along with fast orbital flips (some hundreds of Myrs to pass from $0\degree$ to $180\degree$). Occasionally, the residuals of the classic equilibrium points make $\omega$ oscillate briefly around $\pi/2$ or $3\pi/2$. Such a behaviour is typical of what we obtain when integrating the known distant objects using the parameters given in Tab.~\ref{tab:real} (using the first or the second secular model).
    
\begin{figure}
      \centering
      \includegraphics[width=\textwidth]{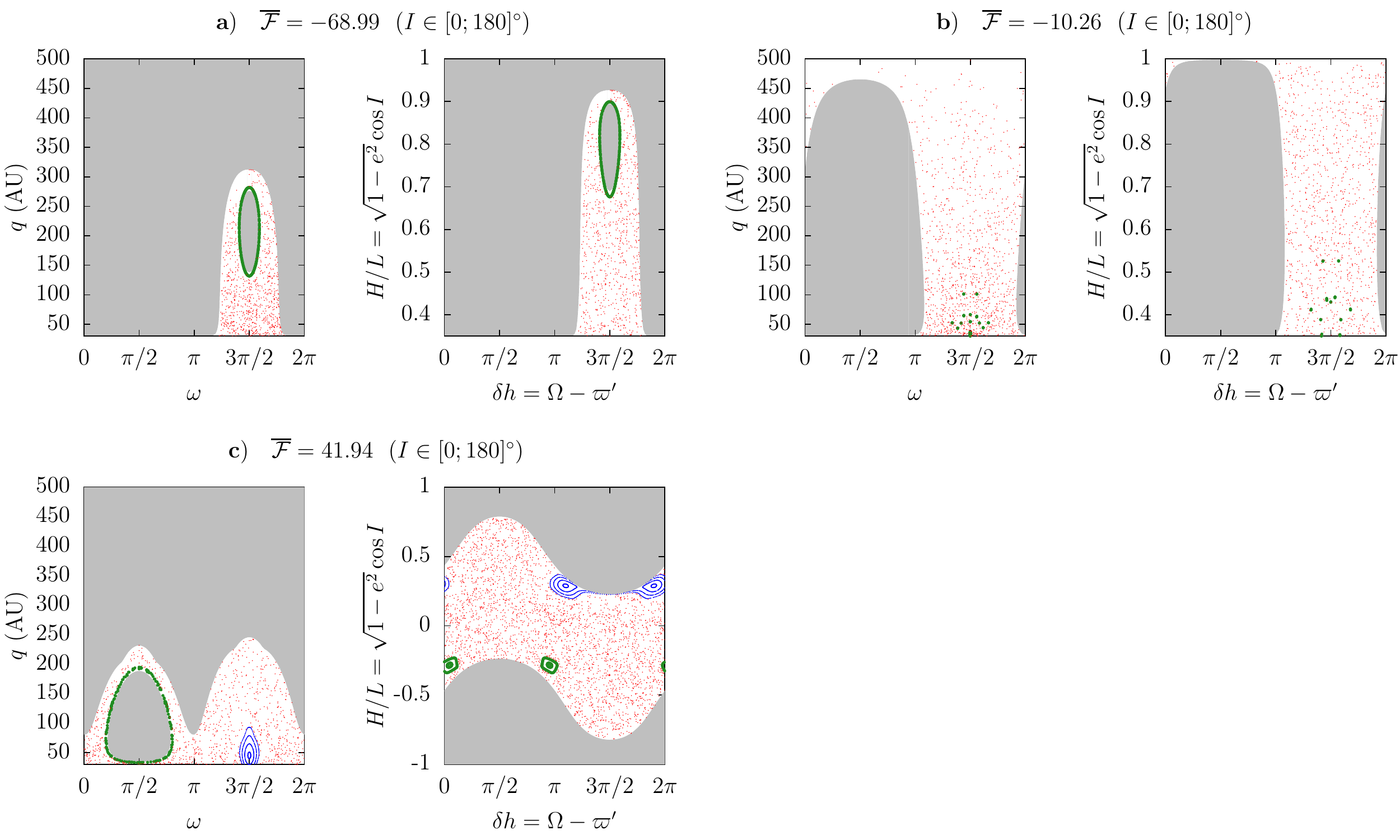}
      \caption{Poincaré maps for a planar perturber. The constant semi-major axis of the particle is $a=500$~AU. Every section in the $(\omega,q)$ plane is made for $\delta h=\pi/2$ and $\dot{\delta h}<0$. Every section in the $(\delta h,H)$ plane is made for $\omega=\pi/2$, with $\dot{\omega}>0$ for (\textbf{a},\textbf{b}) and $\dot{\omega}<0$ for \textbf{c}. The axes ranges are mainly focussed on prograde orbits ($H>0$), but all of the chaotic orbits can flip (see the inclination ranges). On the panel \textbf{c}, a resonance $\omega-\delta h$ (leading to no particular orbital alignment) is visible for $I<90\degree$ in both the surfaces of section. It has a curious two-lobed geometry on the right. In the plane $(\delta h,H)$, a resonance $\Delta\varpi\approx\pi$ is also visible for $I>90\degree$ (it has two islands because $\omega$ and $\delta h$ alternate between oscillation and circulation).}
      \label{fig:P9plan_500}
   \end{figure}

   \bigskip
   For completeness, we also present sections for $a=500$~AU (Fig.~\ref{fig:P9plan_500}). For such high semi-major axes, the phase space is filled with chaos, but the distant planet imposes strong constraints on the shape of the forbidden regions. Hence, even in the chaotic regime, the particle has no other possibility than following temporarily the various $1\!:\!1$ resonances described throughout this section (similar to Fig.~\ref{fig:detail_300}). Some very isolated regular trajectories persists, either as very high-order resonances hidden in the chaotic sea (panel \textbf{b} at $\Delta\varpi\approx 0$), or at very low inclinations (less than $1\degree$). These latter, lying in the very vicinity of the forbidden regions, should persist for all values of the semi-major axis, since we retrieve the integrable model of \cite{BEUST_2016}. On the panel \textbf{c}, small stable regions in the resonances $\omega-\delta h$ and $\omega+\delta h$ are also visible.
   
   \bigskip
   We did not explore orbits with semi-major axes larger than the one of the distant planet. There exist probably other stable equilibrium points for trajectory entirely exterior to its orbit, but the problem becomes too significantly disconnected from the observed Solar System objects.
   
\begin{figure}
      \centering
      \includegraphics[width=\textwidth]{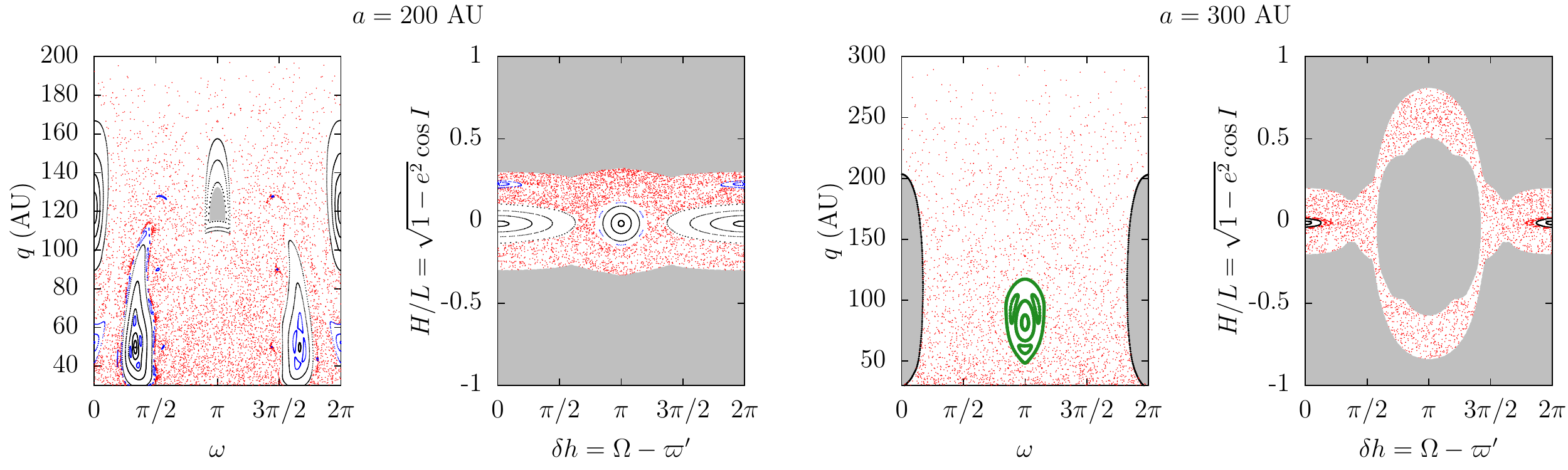}
      \caption{Poincaré maps for a planar perturber. The sections in the $(\omega,q)$ plane are made for $\delta h=0$ and $\dot{\delta h}<0$. The sections in the $(\delta h,H)$ plane are made for $\omega=0$, with $\dot{\omega}<0$. The Hamiltonian value is the same as in Figs.~\ref{fig:P9plan_200}\textbf{f} and \ref{fig:P9plan_300}\textbf{d}. For $a=200$~AU, both the aligned and anti-aligned stable configurations are visible (as well as the classic equilibrium points of $\omega$). For $a=300$~AU, only the aligned one is stable, in a very small region. The section in the $(\omega,q)$ plane features also a resonance $1\!:\!1$ with circulating $\omega$ and $\delta h$ (same as Fig.~\ref{fig:P9plan_300}\textbf{d}).}
      \label{fig:P9plan_Ow0}
   \end{figure}

   All the sections presented above are chosen at $\omega=\pi/2$ and $\delta h=\pi/2$, which does not allow to observe directly the equilibrium configurations with $\omega$ and $\delta h$ both oscillating around $0$ or $\pi$ (they appear only as forbidden zones on Figs.~\ref{fig:P9plan_200}\textbf{f} and \ref{fig:P9plan_300}\textbf{d}). In order to track that kind of behaviour, Fig.~\ref{fig:P9plan_Ow0} shows the previous trajectories projected in sections chosen at $\omega=0$ and $\delta h=0$. These two equilibrium configurations appear from semi-major axes slightly smaller than $200$~AU and become unstable beyond $300$~AU. The aligned configuration is the last to vanish. The instability seems to be due to the oscillation of the perihelion: it goes through the inner planetary region for higher values of $a$, producing a complex pattern of orbit crossings. However, the signature of both configurations remains in the form of sticky chaos, producing in particular high-amplitude orbital flips for anti-aligned orbits (thin corridor available for $a=300$~AU).
   
   \section{Toward a more general case}\label{sec:gen}
   The complete system given by~\eqref{eq:F} has three degrees of freedom, with no possibility to further reduce it without loss of generality. Different strategies can be used to study the dynamics. The most direct one is to perform a series of numerical integrations, in order to compute Lyapunov exponents and stability maps. This would be loosing, however, the benefit of the secular model over the osculation system. Eventually, one can think of a more ``mathematician-like" approach, consisting in the study of an intermediate system, less physically meaningful but still interesting dynamically speaking. This is the strategy adopted in the scope of this paper, by adopting an unrealistic precession rate of $\nu_\omega'=0$. Note that this is a reasonable approximation for small bodies with moderate semi-major axes (say, less than $100$~AU) far from the Kozai equilibrium points (inclination near $63\degree$ or $117\degree$). Indeed the precession rates of such bodies are very fast compared to the outer planet, so that the variation of $\omega'$ can be considered as an adiabatic process (the orbit passes smoothly from one fixed value of $\omega'$ to the other, as long as no major change of topology occurs). However, we will not restrict the study to that region: in return, comparisons with the full secular system will be realised for typical trajectories all along the study.
   
   With the arbitrary use of $\nu_\omega'=0$, the secular system defined by~\eqref{eq:F} has only two degrees of freedom:
   \begin{equation}\label{eq:Fnu0}
      \mathcal{F}(G,H,g,\delta h) = - \nu_\Omega' H + \mathcal{F}_1(G,H,g,\delta h)
   \end{equation}
   and two parameters ($a$ and $\omega'$). Once again, the dynamics can be explored using Poincaré sections. Note that the second angle is this time $\delta h=\Omega-\Omega'$, so one must be careful when comparing with the results from Sect.~\ref{sec:planar}. However, if $\omega'$ is fixed, $\Omega'$ behaves exactly like $\varpi'$, so we can safely identify and link the features from the two cases (equilibrium points, resonances...). In the following, the constant $\omega'$ is chosen to its nominal current value given in Tab.~\ref{tab:orbel}.
   
\begin{figure}
      \centering
      \includegraphics[width=\textwidth]{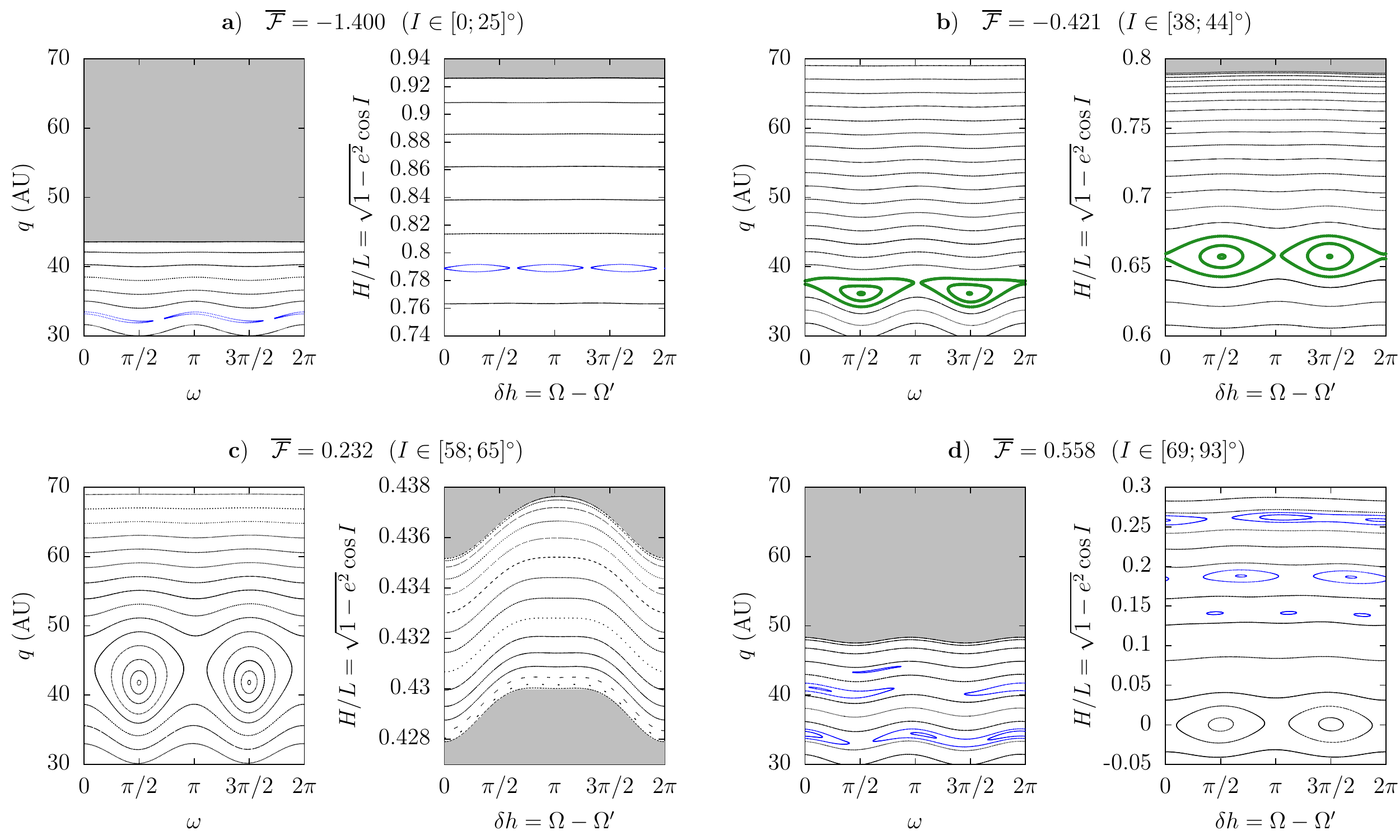}
      \caption{Poincaré maps for a inclined perturber. The constant semi-major axis of the particle is $a=70$~AU. Every section in the $(\omega,q)$ plane is made for $\delta h=\pi/2$ and $\dot{\delta h}<0$. Every section in the $(\delta h,H)$ plane is made for $\omega=\pi/2$, with $\dot{\omega}>0$ for (\textbf{a},\textbf{b}) and $\dot{\omega}<0$ for (\textbf{c},\textbf{d}). The panel \textbf{a} features the resonance $2\omega+3\delta h$. The panels \textbf{b} and \textbf{d} feature the resonance $\omega+\delta h$  and $\omega-\delta h$, respectively, with two different fixed points and horseshoe-type orbits. On the panel \textbf{d}, the resonances $\omega-2\delta h$ and $\omega-3\delta h$ are also visible.}
      \label{fig:P9_70}
   \end{figure}

   \bigskip
   Figure~\ref{fig:P9_70}, computed for $a=70$~AU, shows that even for small semi-major axes, the breaking of symmetry induced by the inclination of the distant perturber produces a richer dynamics, with numerous additional resonances. On the panel \textbf{b} of Fig.~\ref{fig:P9_70}, the two fixed points due to the resonance $\omega+\delta h$ correspond to an alignment between the apsidal line of the small body and the \emph{nodes} line of the distant planet\footnote{On the contrary, apsidal alignment or anti-alignment would have resulted in fixed points on the sections at $\omega'\pm\pi/2$, where $\omega'$ is given in Tab.~\ref{tab:orbel}.}. Hence, the orbital configuration is pretty different from what we obtain for a planar perturber (Fig.~\ref{fig:P9plan_70}), although the same resonance is involved. Moreover, there are this time two fixed points (alignment and anti-alignment) instead of a single one. This holds also for the resonance $\omega-\delta h$ on the panel \textbf{d}. Numerical integrations of the full three-degree-of-freedom secular system reveal that all the libration islands present in Fig.~\ref{fig:P9_70} persist for a precessing $\omega'$. These maps are thus representative of the complete system. Some resonances change critical argument (for instance, $\omega-2\delta h$ turns to $\omega-2\delta h+\omega'$), but not the $1\!:\!1$ one. One can argue that a resonance with critical angle $\omega+\delta h=\varpi-\Omega'$ violates the D'Alembert rules, but as explained below, this resonance is not an artefact due to the fixity of $\omega'$. Actually, the system departs from the three-body problem by its inner component, so one should not be surprised if some ``unusual" resonances show up.
   
\begin{figure}
      \centering
      \includegraphics[width=\textwidth]{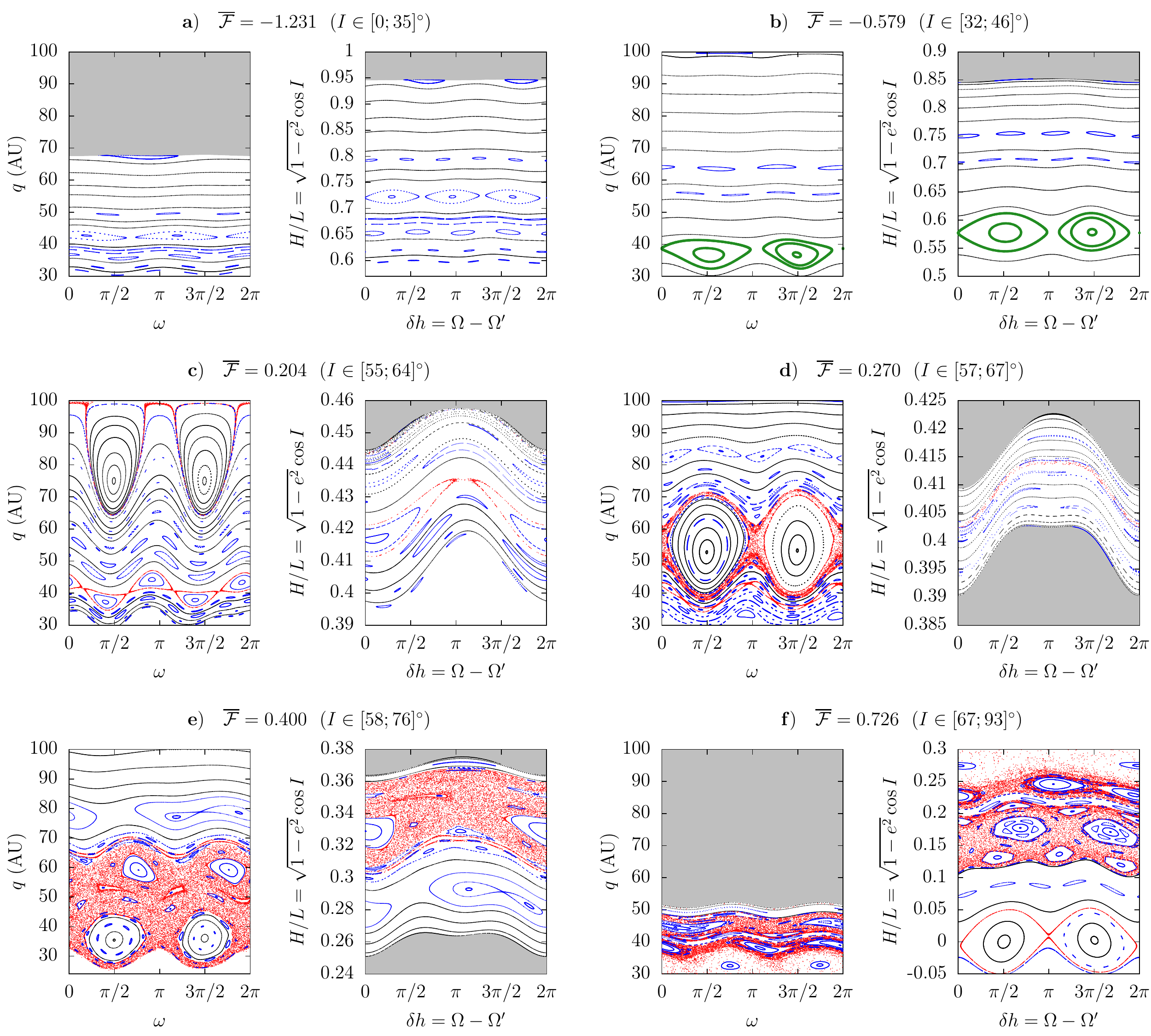}
      \caption{Poincaré maps for a inclined perturber. The constant semi-major axis of the particle is $a=100$~AU. Every section in the $(\omega,q)$ plane is made for $\delta h=\pi/2$ and $\dot{\delta h}<0$. Every section in the $(\delta h,H)$ plane is made for $\omega=\pi/2$, with $\dot{\omega}>0$ for (\textbf{a},\textbf{b},\textbf{c}) and $\dot{\omega}<0$ for (\textbf{d},\textbf{e},\textbf{f}). The panels (\textbf{c},\textbf{d},\textbf{e}) show the shift of the fixed points for $\omega$ at $I\approx 63\degree$ for nearby Hamiltonian values. The surrounding chaotic sea is very wide when the islands are near the semi-major axis of Neptune (panel \textbf{e}). Resonances between $\omega$ and $\delta h$ are very numerous at all ranges of inclination. On the panels \textbf{e} and \textbf{f}, a zoom-in view reveals a very complex pattern of secondary resonances spreading in a fractal-like way.}
      \label{fig:P9_100}
   \end{figure}

   \bigskip
   For $a=100$~AU, a lot of resonances occupy the entire range of inclinations from $0\degree$ to $90\degree$ (Fig.~\ref{fig:P9_100}). As seen on the panel \textbf{e}, the chaotic zone around the two fixed points of $\omega$ is much larger than for a planar perturber (Fig.~\ref{fig:P9plan_100}), allowing large excursions of the perihelion distance. On the panel \textbf{a}, the resonance $\omega+2\delta h$ is visible near the zero-inclination limit. It appears also on the panel \textbf{b}, near the circular orbit, taking the place of the island of apsidal alignment for a planar perturber (compare with Figs.~\ref{fig:P9plan_100} and \ref{fig:P9plan_150}). For a precessing $\omega'$, numerical integrations show that this resonance becomes $\omega+2\delta h-\omega'=\Delta\varpi+\Delta\Omega$, oscillating around $\pi$. Hence, the apsidal alignment $\Delta\varpi = 0$, reported by \cite{BEUST_2016} and present for a planar perturber, does not persist (as such) if the perturber is itself inclined.
   
\begin{figure}
      \centering
      \includegraphics[width=\textwidth]{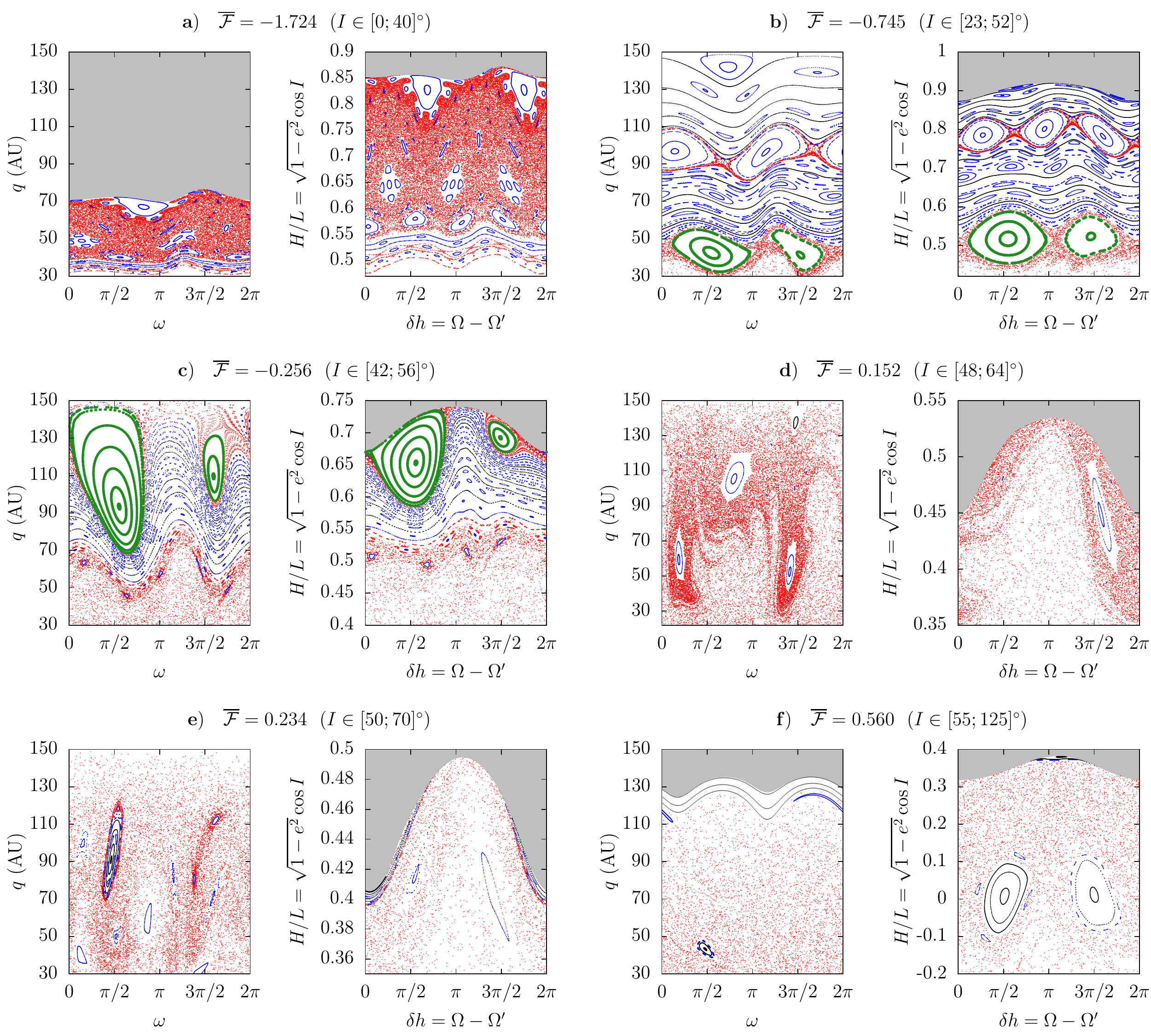}
      \caption{Poincaré maps for a inclined perturber. The constant semi-major axis of the particle is $a=150$~AU. Every section in the $(\omega,q)$ plane is made for $\delta h=\pi/2$ and $\dot{\delta h}<0$. Every section in the $(\delta h,H)$ plane is made for $\omega=\pi/2$, with $\dot{\omega}>0$ for (\textbf{a},\textbf{b},\textbf{c},\textbf{d},\textbf{e}) and $\dot{\omega}<0$ for \textbf{f}.}
      \label{fig:P9_150}
   \end{figure}

   \bigskip
   This is confirmed for larger semi-major axes, since Fig.~\ref{fig:P9_150}, plotted for $a=150$~AU, shows the same $1\!:\!2$ resonance (panels \textbf{a} and \textbf{b}). On the panel \textbf{d}, only the right oscillation island of $\omega$ is visible (at $q\approx 140$~AU); the other is submerged by the chaotic sea. This asymmetry is due to the arbitrary fixed value of $\omega'$. Still on the panel \textbf{d}, the red points form somewhat organised structures around the $3\!:\!1$ resonance, due to sticky chaos. In general, the chaos spreads much faster than for a planar perturber, allowing flips between prograde and retrograde orbits (panel \textbf{f}) even for $a$ as small as $150$~AU (compare with Figs.~\ref{fig:P9plan_150} and \ref{fig:P9plan_200}).
   
\begin{figure}
      \centering
      \includegraphics[width=\textwidth]{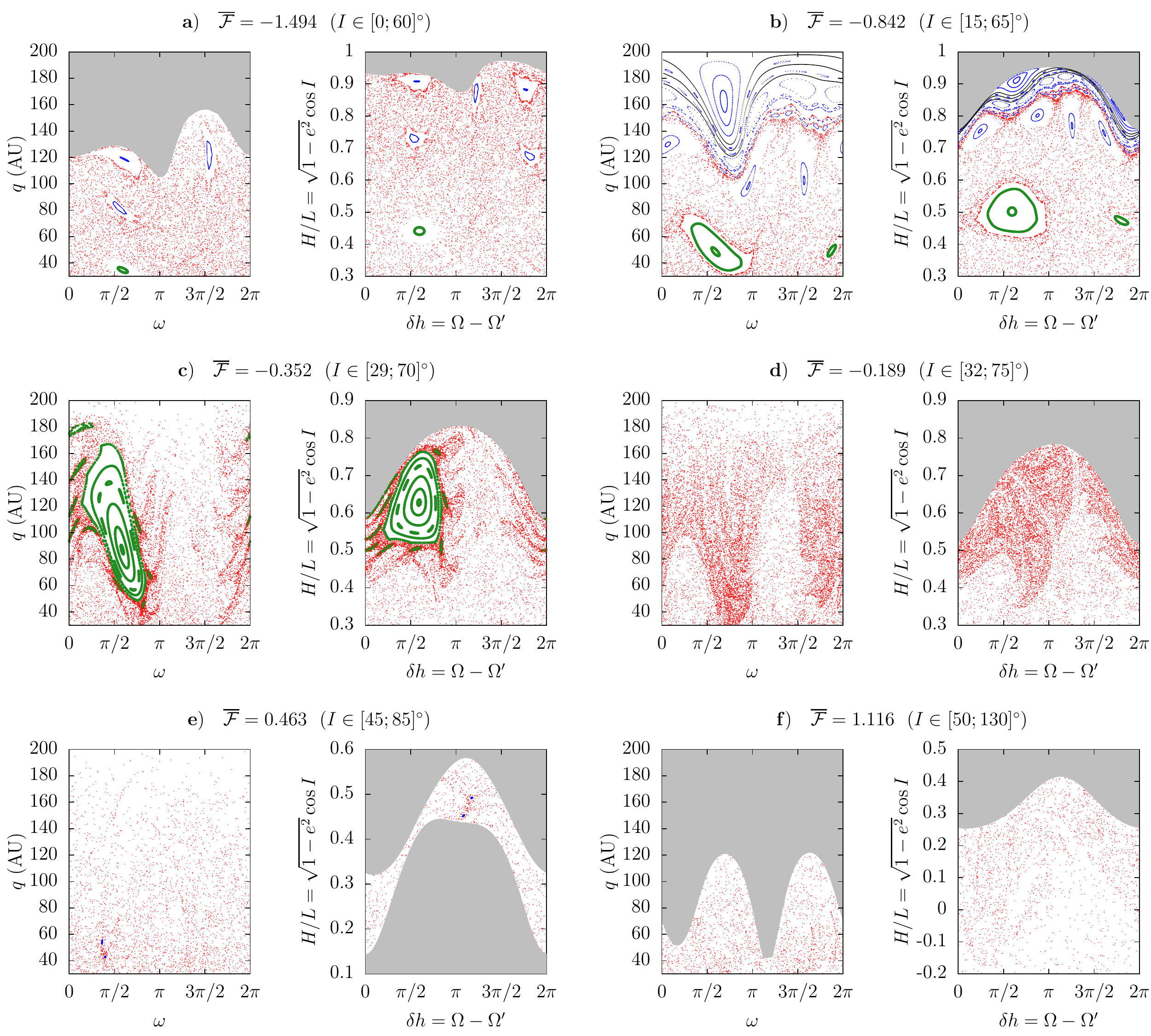}
      \caption{Poincaré maps for a inclined perturber. The constant semi-major axis of the particle is $a=200$~AU. Every section in the $(\omega,q)$ plane is made for $\delta h=\pi/2$ and $\dot{\delta h}<0$. Every section in the $(\delta h,H)$ plane is made for $\omega=\pi/2$, with $\dot{\omega}>0$ for (\textbf{a},\textbf{b},\textbf{c},\textbf{d}) and $\dot{\omega}<0$ for (\textbf{e},\textbf{f}).}
      \label{fig:P9_200}
   \end{figure}

\begin{figure}
      \centering
      \includegraphics[width=\textwidth]{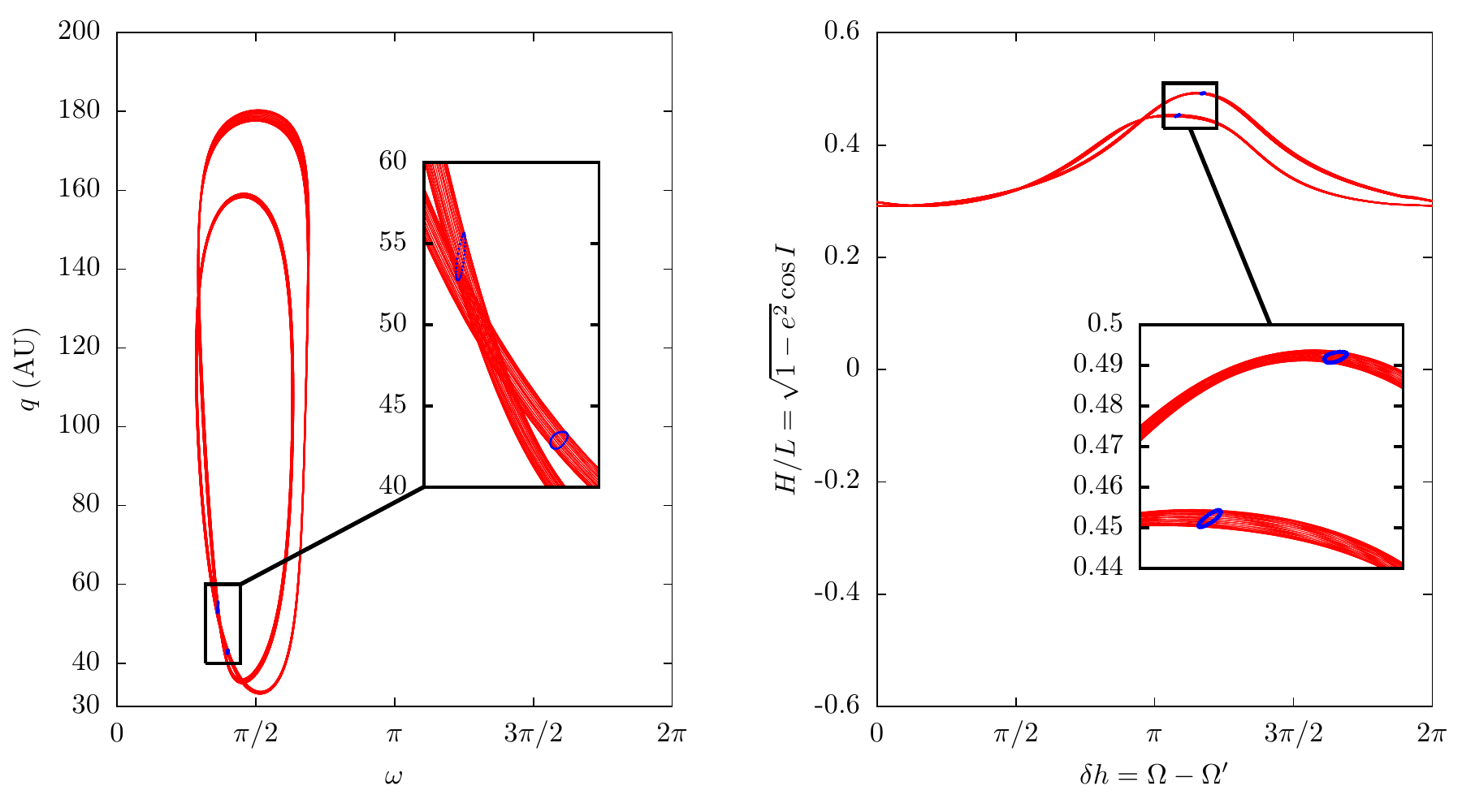}
      \caption{Complete trajectory producing the tiny blue points on the panel \textbf{e} of Fig.~\ref{fig:P9_200} ($a=200$~AU). This is a quasi-periodic orbit featuring the secondary resonance $2\!:\!1$ between $\dot{\delta h}$ and the libration frequency of the resonant angle $1\!:\!1$ between $\dot{\delta h}$ and the libration frequency of $\omega$. On each graph, blue points are added on the two sections (same as Fig.~\ref{fig:P9_200}). Enlarged views show the classic rounded shapes formed by the section crossing points around the periodic trajectory. According to the resonance involved, the fixed points are $2$ on both sides.}
      \label{fig:detail_200}
   \end{figure}

   \bigskip
   For $a=200$~AU, the chaos fills almost all the phase space (Fig.~\ref{fig:P9_200}). The $\omega+2\delta h$ resonance is still present near the zero-inclination limit (panel \textbf{a}) and the circular orbit (panel \textbf{b}). The classic equilibrium points for $\omega$ are almost entirely submerged by the chaotic sea: we found only one quasi-periodic orbit related to oscillations of $\omega$, visible on the panel \textbf{e} (tiny blue points). Fig.~\ref{fig:detail_200} shows this trajectory in details and one can verify that it stays indeed near $I\approx 63\degree$. In that very perturbed case, the sizes of the stability islands do not give any idea of the variations of the orbital elements themselves. In that example, the excursion of the perihelion distance is actually very large, from about $30$ to $180$~AU. Eventually, the only feature still widely emerged in Fig.~\ref{fig:P9_200} is the $1\!:\!1$ resonance between $\omega$ and $\delta h$ (panels \textbf{b} and \textbf{c}). The panel \textbf{d} is drawn for a Hamiltonian value slightly larger than for the panel \textbf{c}, showing the dissolution of the $1\!:\!1$ resonance in chaos. Once again, the finite number of points allows to distinguish structures in the chaotic region, and in particular the general shape of $1\!:\!1$ resonance island. Its signature thus persists, but in the form of sticky chaos. We found similar results in the case of a planar perturber (Fig.~\ref{fig:P9plan_200}, panels \textbf{d}-\textbf{e}).
   
\begin{figure}
      \centering
      \includegraphics[width=\textwidth]{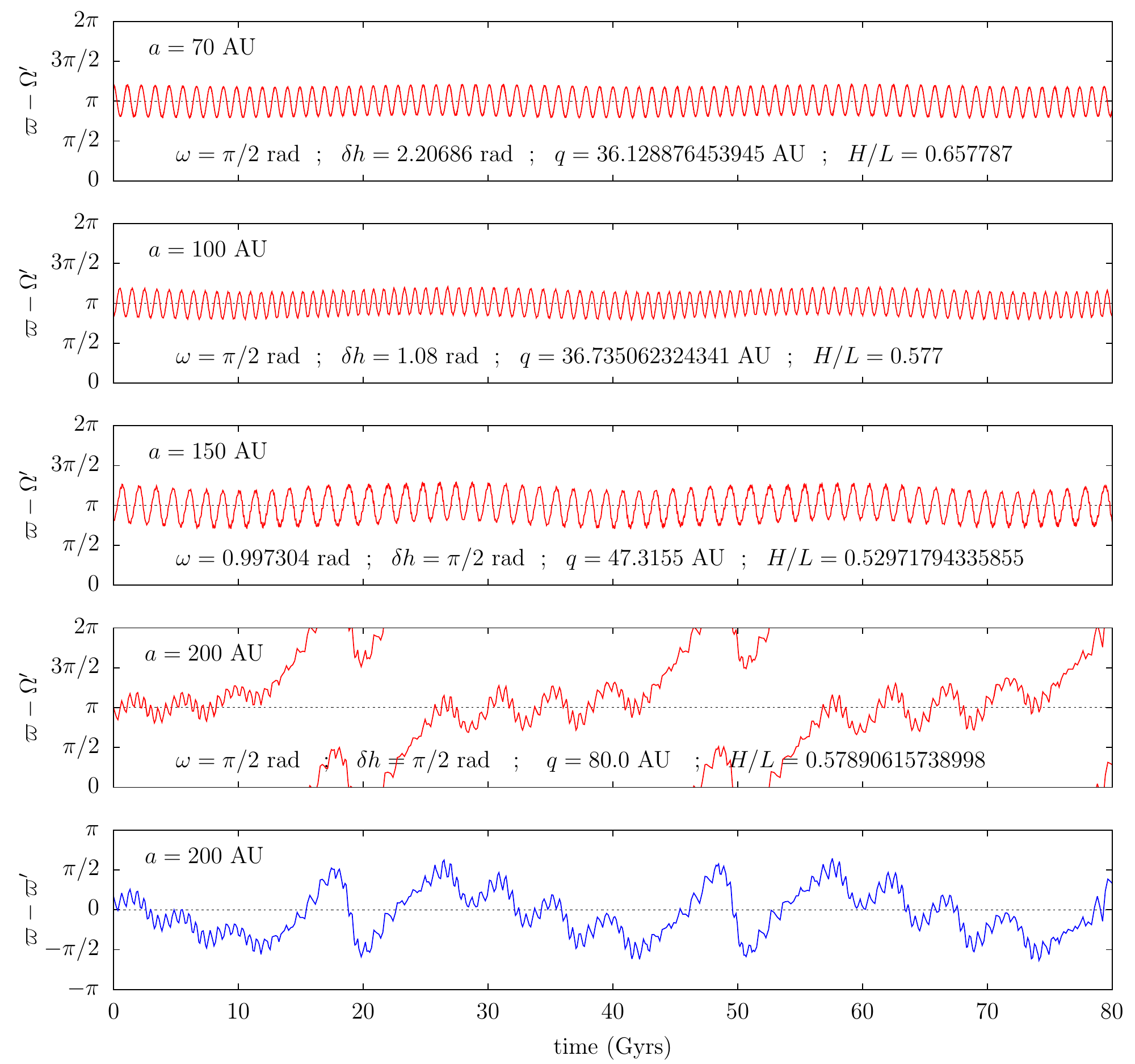}
      \caption{Numerical integrations of the full three-degree-of-freedom secular system. The initial conditions are taken in the $1\!:\!1$ resonance island of the simplified model (inside the left green trajectory in the panel \textbf{b} or \textbf{c} of Figs.~\ref{fig:P9_70}, \ref{fig:P9_100}, \ref{fig:P9_150} and \ref{fig:P9_200}). The details of the initial conditions are written on each graph, and the time evolution of $q$ and $I$ are presented in Fig.~\ref{fig:comp_all_qI}. The new frequency is clearly visible, as a modulation on a $\sim 31$~Gyrs time span, which is the rotation period of $\omega'$. The alignment persists for small semi-major axes ($\varpi-\Omega'\approx\pi$) but is broken beyond some value: for $a=200$~AU, $\varpi-\Omega'$ jumps between $0$ and $\pi$ with the forcing frequency. The lowermost graph, presenting the same trajectory with $a=200$~AU, shows that these jumps produce an overall apsidal alignment ($\Delta\varpi\approx 0$).}
      \label{fig:comp_all}
   \end{figure}

\begin{figure}
      \centering
      \includegraphics[width=\textwidth]{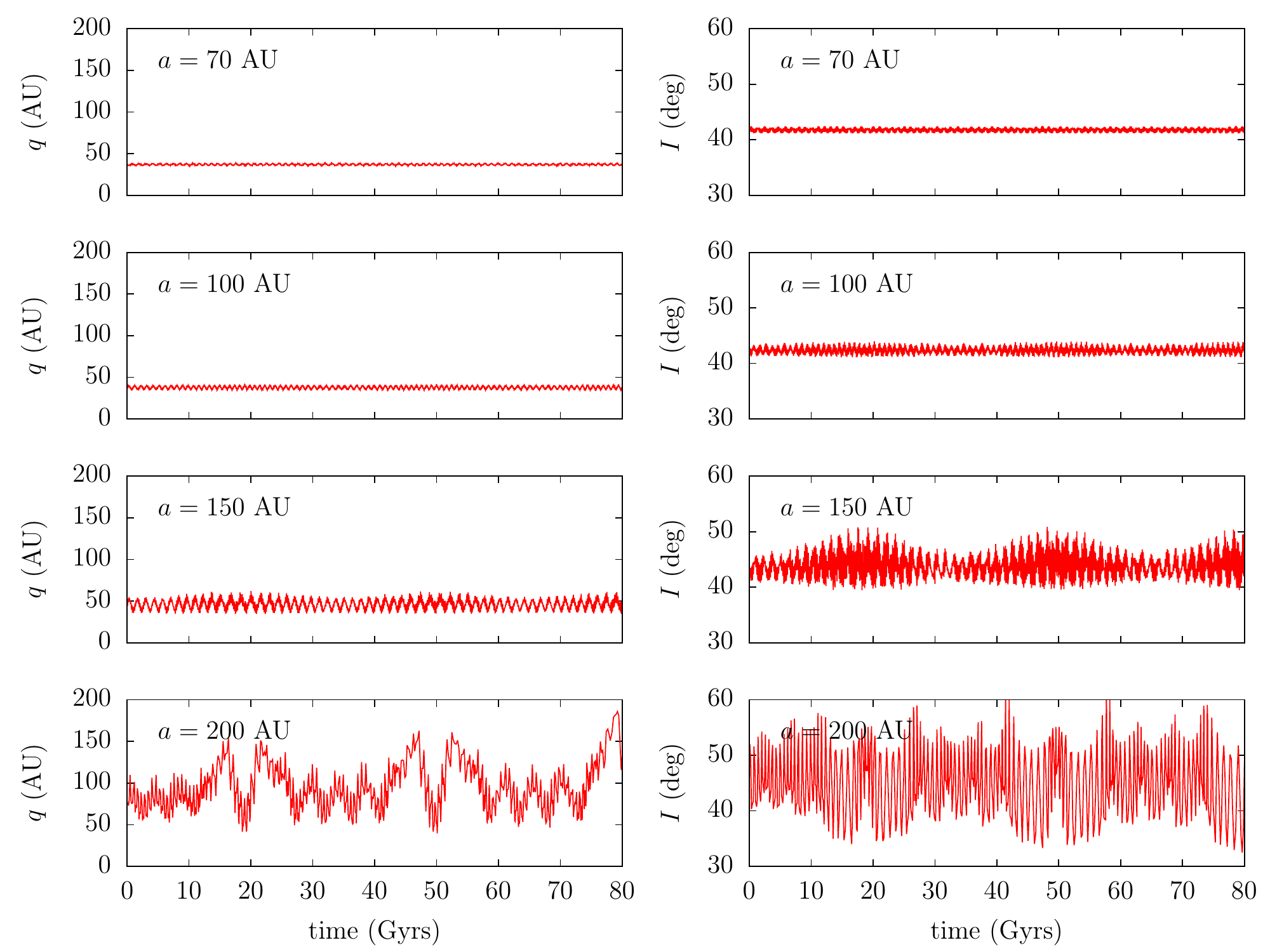}
      \caption{Time evolution of the perihelion distance and of the inclination for the trajectories of Fig.~\ref{fig:comp_all}.}
      \label{fig:comp_all_qI}
   \end{figure}

   \bigskip
   For semi-major axes larger than $200$~AU, our simplified model becomes less relevant and its efficiency to describe the complete system is questionable. Moreover, the exploration of a very chaotic system by the means of Poincaré sections is cumbersome, since there is no guaranty that a stable volume of the phase space does not lie out of the chosen sections. The relevance of the model with $\nu_\omega'=0$ can be assessed by tracking some well-chosen trajectories for increasing values of the semi-major axis. Figures~\ref{fig:comp_all} and \ref{fig:comp_all_qI} present numerical integrations of the complete three-degree-of-freedom system, using initial conditions of trajectories trapped in the $1\!:\!1$ resonance in the simplified model. The model with $\nu_\omega'=0$ proves to be qualitatively relevant from $a=70$ to $150$~AU, since the orbital configuration is conserved (alignment between the apsidal line of the small body and the nodes line of the distant planet). The precession of $\omega'$ adds only extra periodic terms with small amplitudes. For larger semi-major axes, the forcing frequency results in jumps of $\varpi$ from one node of the distant planet to the other ($\varpi-\Omega'=0$ or $\pi$). In that example, these jumps produce an overall \emph{apsidal} alignment, even if the \emph{nodes} were concerned in the first place. This proves that a model with $\nu_\omega'=0$ could be misleading for higher semi-major axes (an extended numerical analysis of the complete secular system would be required, loosing the vantage of a semi-analytical method).
   
\section{Discussion and conclusion}
   We used a secular model to study the long-term dynamics of trans-Neptunian objects perturbed by a distant massive planet. A special attention was given to prograde orbits with perihelion distances beyond Neptune. Classically, the equilibrium points of $\omega$ at $I\approx 63\degree$ or $117\degree$ divide the regions where $\omega$ circulates toward the right and toward the left. The analogous limit for $\Omega$ is $I=90\degree$. This puts some constraints on the regions sensible to a confinement of $\varpi$ in the weakly perturbed case, only possible when $\omega$ and $\Omega$ circulate in opposite directions ($I\in[0;63]\degree$ or $I\in[90;117]\degree$).
   
   For an eccentric planar perturber, the $\Delta\varpi$ resonances reported by \cite{BEUST_2016} are smoothly transported toward non-zero inclinations, both in the aligned and anti-aligned configurations. They are created through $\omega$ and $\Omega$ circulating in opposite directions. Another island of apsidal alignment is highlighted, restricted to inclined orbits, and it has a pretty wide extension for $a>150$~AU. Finally, very particular stable configurations exist from $a\approx 200$ to $300$~AU, in which \emph{both} $\omega$ and $\Omega-\varpi'$ oscillate around $0$ or $\pi$, and $I$ oscillates around $90\degree$. These orbits are aligned or anti-aligned with the distant planet and perpendicular to the planetary plane. They are probably related to the high-inclination aligned population produced in the numerical experiments by \cite{BROWN-BATYGIN_2016}. For higher semi-major axes, these configurations lead the perihelion of the particle inside the inner planetary region, where the successive orbit crossings make them unstable. In addition, the classic equilibrium points of $\omega$ at $\pi/2$ and $3\pi/2$ (for $I\approx 63\degree$) are the source of a chaotic region, spreading all over the phase space when $a$ increases. For $a>300$~AU, the $\Delta\varpi$ resonances are the only remaining stable features, and the chaotic trajectories jump from one of them to the other. Hence, even if there are only small regular regions left beyond $a\sim 300$~AU, the signature of the apsidal alignments and anti-alignments largely remains in the form of sticky chaos. This contributes probably to a large extent to the aligned bodies coming from the simulations by \cite{BATYGIN-BROWN_2016}.
   
   The model with an inclined perturber with a fixed argument of perihelion, even if not strictly realistic, gives an idea of the secular dynamics in the general case. As before, the $1\!:\!1$ resonance between the two degrees of freedom is the most persistent structure. For small semi-major axes (say below $150$~AU), it results in an unusual alignment between the apsidal line of the small body and the nodes line of the distant planet, which persists in the unsimplified three-degree-of-freedom secular system. In a large sample, though, its signature is probably unnoticeable, mixed up with the numerous other features. As expected, the chaos spreads faster than for a planar perturber. In the unsimplified secular system, the extra forcing frequency adds even more chaos in the system: in practice, the distant perturber can be neglected only for very small semi-major axes (say below $70$~AU). For $a=150$~AU already, only a small portion of the phase space is still filled with regular trajectories and large orbital flips become possible (switch between prograde and retrograde orbits). An inclined distant perturber should thus imply a substantial amount of retrograde objects with $a>150$~AU, without even mentioning close encounters. Such objects are indeed observed, as reported by \cite{BATYGIN-BROWN_2016b}, even though none has been observed yet with a perihelion beyond the semi-major axis of Neptune. Finally, we did not observe equilibrium points for $\Delta\Omega$ as in the simple analytical model by \cite{BATYGIN-BROWN_2016} (see their Fig.~7). This means that their effect can persist only through sticky chaos in our more general model.
   
   We conclude that even in the secular system, which is free from any diffusion of semi-major axis, the orbital alignment of distant objects (as well as any organised structure beyond $a\sim 200$~AU) induced by a distant perturber are almost only produced through a collective behaviour of chaotic trajectories, each of them spending more time in preferred locations of the phase space but still wandering ``everywhere". In particular, this is the case of the six objects with $a>250$~AU analysed by \cite{BATYGIN-BROWN_2016}. Hence, the fact that the seemingly clustered objects do not remain efficiently shepherded in long-term numerical simulations is not a sufficient argument to rule out the hypothesis of an external perturber in the Solar System, neither the discovery of distant bodies \emph{out} of the accumulation regions. For instance, the recently discovered trans-Neptunian object 2015\,GT$_{50}$ \citep{BANNISTER-etal_2017} is neither aligned nor anti-aligned: according to the MPC database, in this case $\Delta\varpi\approx 272\degree$ using for the distant perturber the data of Tab.~\ref{tab:orbel}. Here, the only viable approach is to deal with distributions, which should be peaked or not around preferential configurations. Whereas it would be simple (in theory) to confirm the existence of a distant planet by direct observation, to exclude it requires the analysis of a large observational sample, pretty hard to obtain.
   
   This study was limited to secular orbits away from mean-motion resonances with the planets. Mean-motion resonances are a well-known source of orbital confinements, and they are likely to allow a lot of stable configurations. However, note that particles with slowly diffusing semi-major axes, even with successive transient captures in various mean-motion resonances, would present in overall the signature of a \emph{non-resonant} secular dynamics. Hence, in order to allow specific confinements, such resonance trappings must be long enough. Since a long-term resonant interaction with Neptune cannot confine $\omega$ in a manner that fits the orbital distribution of the observed distant objects \citep[][]{SAILLENFEST-etal_2016b}, the mean-motion resonances involved should be with the hypothetical distant planet. That aspect was studied in particular by \cite{DELAFUENTEMARCOS-DELAFUENTEMARCOS_2016}. An averaged model in the resonant case could also be developed (by taking advantage of the separation between the timescales involved), but the corresponding Poincaré maps are expected to be extremely rich and probably difficult to interpret.
   
\begin{acknowledgements}
   We thank two anonymous referees who helped us to improve the paper. This work was partly funded by Paris Sciences et Lettres (PSL).
\end{acknowledgements}

\bibliographystyle{aps-nameyear}
\bibliography{CM_2016b}

\appendix
\newpage
\section{Secular Hamiltonian in the completely planar case}\label{Asec:plan}
   \begin{figure}[h]
      \centering
      \includegraphics[width=\textwidth]{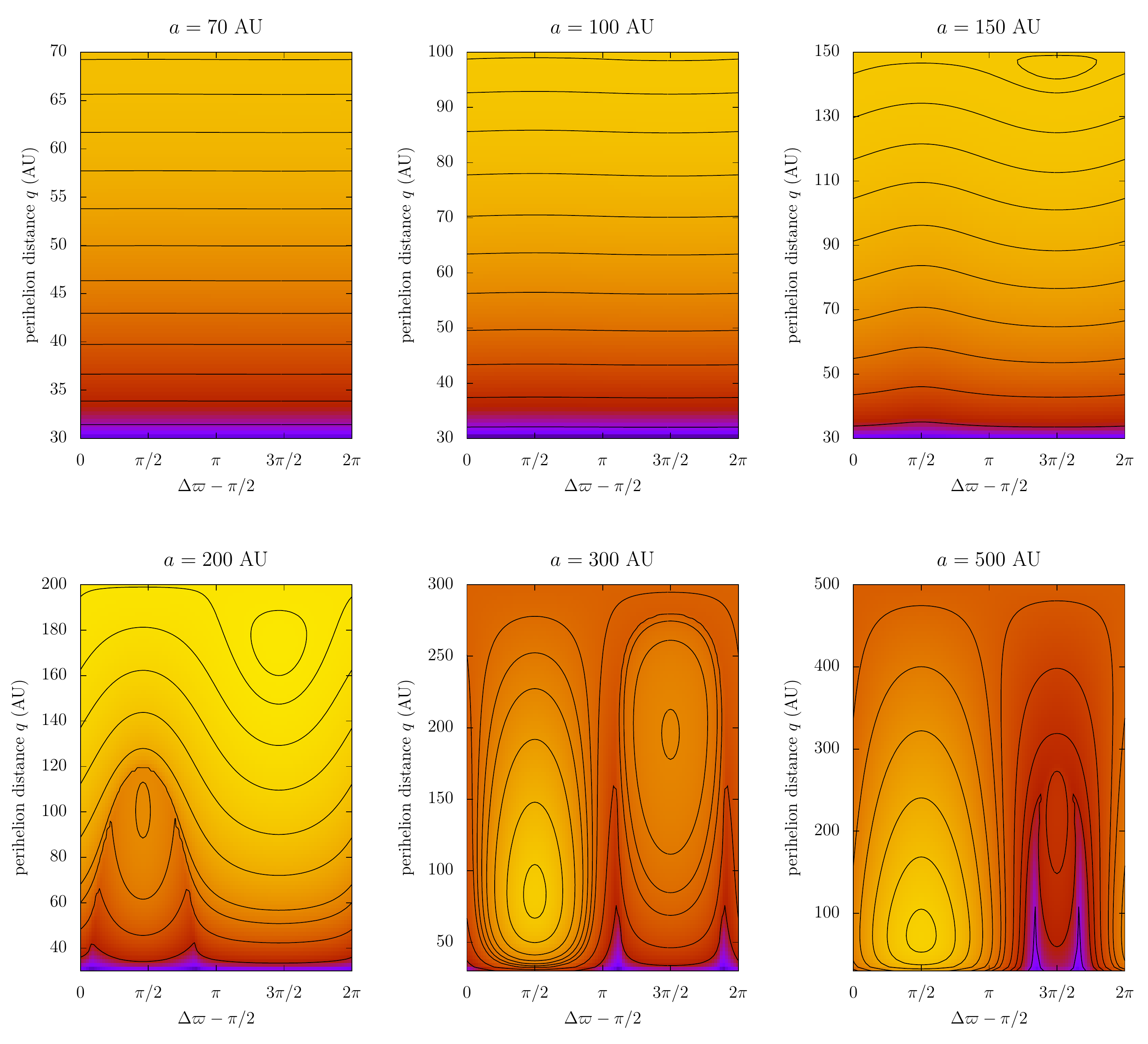}
      \caption{Level curves of the secular Hamiltonian in the completely planar case \citep[same as][]{BEUST_2016}. The perturbations of both the internal planets and the distant one are completely taken into account (numerical average). In order to ease the comparison with the Poincaré sections throughout this article, we use the perihelion distance $q$ instead of $e$, and $\Delta\varpi-\pi/2$ instead of $\Delta\varpi$.}
      \label{fig:Hsec-plan}
   \end{figure}
  
\newpage 
\section{Initial conditions of observed distant objects}\label{Asec:ic}
   \begin{table}[h]
      \centering
      \begin{tabular}{l || c | c | c | c || c | c || c | c}
         name & $a$ & $q$ & $H/L$ & $\omega$ & $\Omega-\varpi'$ & $\overline{\mathcal{F}}_1$ & $\Omega-\Omega'$ & $\overline{\mathcal{F}}_2$ \\
         & (AU) & (AU) & & (rad) & (rad) & & (rad) & \\
         \hline
         2012\,VP$_{113}$ & $255.9$ & $80.54$ & $0.6650$ & $5.131$ & $3.277$ & $-1.561$ & $5.895$ & $-3.062$ \\
         2004\,VN$_{112}$ & $316.4$ & $47.32$ & $0.4745$ & $5.708$ & $2.845$ & $0.672$ & $5.463$ & $1.155$ \\
         2013\,RF$_{98}$  & $349.2$ & $36.09$ & $0.3851$ & $5.441$ & $2.873$ & $3.383$ & $5.491$ & $0.795$ \\
         2010\,GB$_{174}$ & $367.1$ & $48.79$ & $0.4633$ & $6.071$ & $3.974$ & $8.277$ & $0.308$ & $8.715$ \\
         2007\,TG$_{422}$ & $476.5$ & $35.57$ & $0.3593$ & $4.986$ & $3.664$ & $36.11$ & $6.282$ & $35.80$ \\
         Sedna            & $493.1$ & $76.03$ & $0.5219$ & $5.438$ & $4.215$ & $52.82$ & $0.550$ & $57.43$
      \end{tabular}
      \caption{Heliocentric osculating elements at current time of the six objects with $a>250$~AU used by \cite{BATYGIN-BROWN_2016}. These elements are computed using AstDyS database (\texttt{http://hamilton.dm.unipi.it/astdys/}), where the value of $\varpi'$ and $\Omega'$ are taken from Tab.~\ref{tab:orbel}. On the right, the corresponding value of the secular Hamiltonian is given using the osculating elements as an approximation of the secular ones. The two models considered here are written ``1" for the planar perturber and ``2" for the inclined perturber with $\nu_\omega'=0.$}
      \label{tab:real}
   \end{table}

\end{document}